\documentclass[prd,superscriptaddress,nofootinbib,notitlepage]{revtex4-1}
\usepackage{epsfig,amsmath,amssymb,slashed}
\usepackage{graphicx}
\usepackage{color}
\usepackage{siunitx}
\usepackage{color}
\usepackage{bm}
\usepackage{subcaption}
\usepackage{tabularx}
\usepackage{float}
\newcommand{\bra}[1]{\langle #1|}
\newcommand{\ket}[1]{|#1\rangle}

\newcommand{\di}{{\rm d}}
\newcommand{\ii}{i}

 \def\wT{{\widehat T}}

\def\wpsi{{\widehat{\psi}}}
 
\def\wrho{{\widehat{\rho}}}

\def\wa{\widehat a}
\def\wad{\widehat a^{\dagger}}

\def\huno{{\mathrm{H^{(1)}_\nu}}}
\def\hdue{{\mathrm{H^{(2)}_\nu}}}
\def\hunopiu{{\mathrm{H^{(1)}_{\nu+1}}}}
\def\hduepiu{{\mathrm{H^{(2)}_{\nu+1}}}}
\def\hunoC{{\mathrm{H^{(1)}_{i\mu}}}}
\def\hdueC{{\mathrm{H^{(2)}_{i\mu}}}}

\newcommand{\Tr}{{\rm Tr}}  
  
\newcommand{\e}{{\rm e}}
\newcommand{\kk}{{\rm k}}

\newcommand{\x}{{\rm x}}

\newcommand{\be}{\begin{equation}}
\newcommand{\ee}{\end{equation}}                                                                               
\def\bea{\begin{eqnarray}}
\def\eea{\end{eqnarray}}      
\begin{document}

\title{Negative pressure as a quantum effect in free-streaming in the cosmological background} 

\author{F. Becattini, D. Roselli}
\affiliation{Universit\`a di Firenze and INFN Sezione di Firenze, Florence, Italy}

\begin{abstract}
We present a study of energy density and pressure of a free real scalar quantum field after its 
decoupling from a thermal bath in the spatially flat Friedman-Lema\^itre-Robertson-Walker space-time 
by solving the Klein-Gordon equation both analytically and numerically for different predetermined 
scale factor functions $a(t)$. The energy density and pressure, defined by subtracting 
the vacuum expectation values at the decoupling time, feature corrections with respect to the classical 
free-streaming solution of the relativistic Boltzmann equation. We show that if the expansion rate is 
comparable or larger than $mc^2/\hbar$ or $KT_0/\hbar$ where $m$ is the mass and $T_0$ the decoupling 
temperature, both energy density and pressure gets strong quantum corrections which substantially modify 
their classical dependence on the scale factor $a(t)$ and drive pressure to large negative values. 
For a minimally coupled field with a very low mass in an expanding de Sitter universe quantum 
corrections are dominant driving pressure and energy density to become asymptotically constant with 
an equation of state $p/\varepsilon \simeq -1$, thereby mimicking a cosmological constant. 
For a minimally coupled massless field, quantum corrections are asymptotically dominant for any
accelerated expansion. 
\end{abstract}

\maketitle

\section{Introduction}
\label{sec:intro}

In a previous paper \cite{becarose1} we determined the analytic expression of the energy 
density and pressure of a real scalar field after its decoupling from a thermal bath in the 
cosmological background, defined by the spatially flat Friedman-Lema\^itre-Robertson-Walker 
(FLRW) metric:
\begin{equation}\label{FlatRWF}
    \di s^2=\di t^2-a^2(t)\left(\di x^2+\di y^2+\di z^2\right).
\end{equation}
where $a(t)$ is the scale factor.
The energy density and pressure were defined as those induced by the excitations with respect to the
vacuum state $\ket{0_{t_0}}$ at the decoupling time, i.e. the lowest lying state of the Hamiltonian 
at the decoupling time $t_0$. In formulae, by defining the vacuum-subtracted stress-energy tensor 
as:\footnote {It is very interesting to observe that by defining:
$$
 \wT^{\mu\nu}_{\rm ren} = \wT^{\mu\nu} -  \bra{0_{t_0}} \wT^{\mu\nu}(x) \ket{0_{t_0}}
$$
this operator fulfills the first three Wald's postulates \cite{wald1977} which characterize a renormalized
stress-energy tensor with desired features.}
\begin{equation}\label{renset}
    \langle \wT^{\mu\nu}(x) \rangle =\Tr\left(\wrho(t_0)\wT^{\mu\nu}(x) \right)-
    \bra{0_{t_0}} \wT^{\mu\nu}(x) \ket{0_{t_0}} =
    (\epsilon + p) u^\mu u^\nu - p g^{\mu\nu}
\end{equation}
where $u^\mu=(1,0,0,0)$ in the coordinates $(t,{\bf x})$ and $\wrho(t_0)$ is the local thermodynamic 
equilibrium density operator:

\be\label{densop}
 \wrho(t_0) = \dfrac{1}{Z} \exp\left[ -\widehat H(t_0)/T(t_0) \right] = 
  \dfrac{1}{Z} \exp\left[ -\int \di^3 \x \; \wT^{00}(x) \right],
\ee
where we have set $a(t_0)=1$, what will be understood throughout the paper.

The definition \eqref{renset}, in the language of quantum statistical mechanics, corresponds to the 
stress-energy tensor of classical relativistic particles decoupling from the cosmological thermal bath 
at the time $t_0$ and freely-falling thereafter, whose expression, in terms of energy density and 
pressure can be calculated by means of relativistic kinetic theory and reads \cite{kremer,Kolb}:
\begin{equation}\label{classical}
\begin{split}
\varepsilon(t)&=\dfrac{1}{\left(2\pi\right)^3a^4(t)}\int \di^3\kk \; \omega_k(t)
 n_B\left(\dfrac{\omega_k(t_0)}{T(t_0)}\right),\\
 p(t)&=\dfrac{1}{\left(2\pi\right)^3a^4(t)}\int \di^3\kk \; \dfrac{\kk^2}{3\omega_k(t)}
 n_B\left(\dfrac{\omega_k(t_0)}{T(t_0)}\right),   
\end{split}
\end{equation}
where
$$
 n_B(x) = \dfrac{1}{\e^x -1} \qquad  {\rm and} \qquad
 \omega_k(t)=\sqrt{\kk^2+m^2a^2(t)},
$$
and ${\bf k}$ denotes the conserved covariant components of the four-momentum. In fact, the
subtraction of the vacuum expectation value at $t_0$ in the definition \eqref{renset} is
equivalent to the familiar normal ordering of the creation and annihilation operators at $t_0$
so that for $T(t_0)=0$ the stress energy tensor vanishes. Furthermore, the stress-energy 
tensor \eqref{renset} fulfills the continuity equation $\nabla_\mu \langle \wT^{\mu\nu}(x) 
\rangle = 0$ because both the density operator and the vacuum state are at some fixed time, so
it is compatible with the General Relativity gravitational field equation. 

The full calculation of \eqref{renset} by using quantum field theory in curved space-time has 
been carried out in ref.~\cite{becarose1}. Denoting by $\wpsi(x)$ the field operator fulfilling 
the Klein-Gordon equation in the FLRW metric:
\be\label{kg}
 ( \Box + m^2 - \xi R ) \wpsi = 0
\ee
with $\xi$ the adimensional parameter coupling the field to the curvature, we can expand it
as a superposition of covariant momentum components:
\be\label{field}
 \wpsi(x)  = \dfrac{1}{a(t)(2\pi)^{3/2}} \int \di^3 \kk \;\left( v_{k}(t) \e^{\ii {\bf k}\cdot{\bf x}}
 \wa{({\bf k})} + v^*_{k}(t) \e^{-\ii {\bf k}\cdot{\bf x}} \wad{({\bf k})}\right)
\ee
The mode functions $v_k$ are the solutions of the Klein-Gordon equation which make the vacuum of
the Hamiltonian at the time $t_0$ coinciding with the vacuum of the field annihilated by the
operators $\wa{({\bf k})}$. The $v_k$ fulfill the equation \cite{mukhanov,birrell}:
\begin{equation}\label{kg2}
    v''_k(\eta)+ \Omega^2_k v_k(\eta)=0,
\end{equation}
in the conformal time $\eta= \int^t_{t_0} \di t' (1/a(t'))$, with:
\be\label{Omega}
\Omega^2_k = \kk^2+m^2a^2(\eta)-\left(1-6\xi\right)\dfrac{a''(\eta)}{a(\eta)}
\ee
Note that for the massless case the equation \eqref{kg2} becomes:
\begin{equation*}
    v''_k+\left(\kk^2+(6\xi-1)\frac{a''}{a} \right) v_k=0,
\end{equation*}
which, for $\xi=1/6$, is the same equation for the modes of a massless field in a flat space-time. 
Instead, for $\xi>0$ the term $(6\xi-1)a''/a$ acts as an effective time-dependent mass term.

The initial conditions are dictated by the choice of the vacuum:
\begin{equation}\label{condin}
v_k(0)=\dfrac{1}{\sqrt{2\omega_\xi\left(0,k\right)}}\quad 
\left.\dfrac{\di v_k}{\di \eta}\right|_{\eta=0}=-\dfrac{i}{2v_k(0)}+\left(1-6\xi\right)a'(0)v_k(0).
\end{equation}
with:
\be\label{omegaxi}
 \omega_\xi\left(\eta,k\right)=\sqrt{\kk^2+m^2a^2(\eta)+6\xi\left(1-6\xi\right)\dfrac{a'^2(\eta)}{a^2(\eta)}}.
\ee
and the normalization condition
\begin{equation}\label{wronskian}
    W\left[v_k,v^*_k\right]=v_k(\eta)v'^*_k(\eta)-v'_k(\eta)v^*_k(\eta)=i.
\end{equation}
which makes the commutation relations the standard ones:
$$
 [\wa{({\bf k})},\wad{({\bf k}')}] = \delta^3({\bf k}-{\bf k}')
$$
For $\xi\in\left[0,1/6\right]$ the effective Hamiltonian is bounded from below at every time, that is it 
is always possible to find a minimum eigenvalue and define the proper initial conditions for the modes 
associated with the state of local thermodynamic equilibrium at the decoupling. Note that the most
significant values for $\xi$ are $\xi=0$, corresponding to the minimal coupling, and $\xi=1/6$ corresponding 
to the conformal coupling. 

The obtained expressions of the energy density and pressure from the definition \eqref{renset} read:
\begin{equation}\label{epsilonp}
\begin{split}
a^4(\eta)\varepsilon(\eta)&=\dfrac{1}{\left(2\pi\right)^3}
\int\di^3\kk\; \omega_k(\eta)K_k(\eta)\, n_B\left(\dfrac{\omega_\xi\left(0,k\right)}{T(0)}\right),\\
a^4(\eta)p(\eta)&=\dfrac{1}{\left(2\pi\right)^3}\int\di^3\kk\; \omega_k(\eta)\Gamma_k(\eta) \,
n_B\left(\dfrac{\omega_\xi\left(0,k\right)}{T(0)}\right),
\end{split}
\end{equation}
where the functions $K_k$ and $\Gamma_k$ are given by
\begin{equation}\label{funzKGAMMA}
    \begin{split}
K_k(\eta)&=\dfrac{1}{\omega_k(\eta)}\left[\left|v'_k(\eta)\right|^2+\omega^2_k(\eta)
\left|v_k(\eta)\right|^2-\left(1-6\xi\right)\dfrac{a'(\eta)}{a(\eta)}
\left(v'_k(\eta)v^*_k(\eta)+v'^*_k(\eta)v_k(\eta)\right)\right],\\
\Gamma_k(\eta)&=\dfrac{1}{\omega_k(\eta)}\left[\left(1-4\xi\right)\left|v'_k(\eta)
\right|^2+\dfrac{\gamma_k(\eta)}{3}\left|v_k(\eta)\right|^2-\left(1-6\xi\right)
\dfrac{a'(\eta)}{a(\eta)}\left(v'_k(\eta)v^*_k(\eta)+v'^*_k(\eta)v_k(\eta)\right)\right],
    \end{split}
\end{equation}
and $\omega_k(\eta), \gamma_k(\eta)$ read:
\begin{equation}\label{funzOmega}
    \begin{split}
\omega_k(\eta)&=\sqrt{\kk^2+m^2a^2(\eta)+\left(1-6\xi\right)\dfrac{a'^2(\eta)}{a^2(\eta)}},\\
\gamma_k(\eta)&=\left(12\xi-1\right)\kk^2+3\left(4\xi-1\right)m^2a^2(\eta)+3\left(1-6\xi\right)
\dfrac{a'^2(\eta)}{a^2(\eta)}-12\xi\left(1-6\xi\right)\dfrac{a''(\eta)}{a(\eta)},
    \end{split}
\end{equation}
The equations \eqref{epsilonp} come down to the classical free-streaming \eqref{classical} in flat 
space-time and, likewise, for $T_0 = 0$ they provide vanishing energy density and pressure
so that the definition \eqref{renset} can be considered as the thermal part of the stress-energy 
tensor at the decoupling. However, they involve corrections to the \eqref{classical} which are 
owing to the quantum nature of matter and are encoded in the mode functions $v_k$. 
For $\xi=0,1/6$ such quantum corrections vanish at the decoupling time $\eta=0$ and the energy 
density and pressure get the classical relativistic thermodynamic form \eqref{classical} 
(see also \cite{capozziello}). Yet, in general, they become non-vanishing thereafter \cite{becarose1} 
as we will see later on. Recently, quantum corrections to the free streaming have been found to 
generally occur in flat spacetime as well for a general density operator \cite{tinti}.
A remarkable exception is that of conformally invariant theories, such as the case with $\xi=1/6, 
m=0$ free scalar field (and the electromagnetic field as well) for which quantum corrections vanish 
altogether, as specifically shown in ref. \cite{becarose1}.

Interestingly, the equations \eqref{epsilonp} imply the following inequalities \cite{becarose1}:
\be\label{wlimits}
 \begin{split}
   \varepsilon > 0 & \qquad -\frac{1}{3} \varepsilon < p < \frac{1}{3} \varepsilon 
   \qquad \xi = \frac{1}{6} \\
   \varepsilon > 0 & \qquad - \varepsilon < p < \varepsilon \qquad\qquad \xi = 0
\end{split}   
\ee
and, for $\xi=0$ and $m=0$:
$$
     p > - \frac{\varepsilon}{3}     
$$
Indeed, there is no general constraint on the positivity of pressure after the decoupling. Nevertheless,
they fulfill the {\em dominant energy condition}, which stipulates that for any time-like future-oriented
vector field $V^\mu$, the vector field $T^{\mu\nu}V_\nu$ is in turn time-like future-oriented and it
ensures causality in energy propagation. For a stress-energy tensor of the perfect-fluid form 
\eqref{renset}, such condition indeed implies $\varepsilon > 0$ and $|p| < \varepsilon$.

In this work we carry out a calculation of $\varepsilon$ and $p$ for different models of the expansion, 
that is for different given scale factor functions $a(t)$. We will show that, whenever the expansion rate 
$\dot a/a$ is comparable or larger than $mc^2/\hbar$ and $KT_0/\hbar$ where $m$ is the mass and $T_0$ 
the decoupling temperature, energy density and pressure get large quantum corrections with respect to
their classical values \eqref{classical} and pressure turns negative. We will show how the full quantum 
solution of the free-streaming problem, under the above conditions, implies a substantial change of the 
dependence of energy density and pressure on the scale factor and of the ratio $p/\varepsilon$ which 
can become relatively large and negative, within the aforementioned bounds. It should be pointed out 
that the purpose of this work is, in the first place, to study the problem of quantum field effects on 
free-streaming in a cosmological background and not to model the actual expansion of the physical universe.

\subsection*{Notation}

In this paper we use the natural units, with $\hbar=c=K=1$. The Minkowskian metric tensor $\eta$ is 
${\rm diag}(1,-1,-1,-1)$; for the Levi-Civita symbol we use the convention $\epsilon^{0123}=1$.\\ 
We will use the relativistic notation with repeated indices assumed to be saturated. Operators in Hilbert 
space will be denoted by a wide upper hat, e.g. $\widehat H$. Sometimes, scalar products of four-vectors
such as $V_\mu U^\mu$ will be denoted with a dot, i.e. $V \cdot U$. The Riemann tensor is defined as
$R^\mu_{\;\; \rho\lambda\nu} = \partial_\lambda \Gamma^\mu_{\rho\nu}- \partial_\nu \Gamma^\mu_{\rho\lambda}
+ \Gamma^\mu_{\lambda\sigma} \Gamma^\sigma_{\rho\nu}- \Gamma^\mu_{\nu\sigma} \Gamma^\sigma_{\rho\lambda}$

\section{General considerations}
\label{sec:gen}

In order to reckon the quantum field effects on energy density and pressure after freeze-out, it 
is useful to write them down by using the cosmological time $t$ instead of the conformal time $\eta$ and 
the mode functions $u_k(t) \equiv v_k(t)/a(t)$. We will confine to the case of $\xi=0$, which is 
simpler for illustration purposes. In this case, the energy density and pressure \eqref{epsilonp} read:
\begin{equation}\label{epsilonpt}
\begin{split}
\varepsilon(t)&= \dfrac{1}{\left(2\pi\right)^3}\int\di^3\kk \; \left[ |\dot u_k(t)|^2 + 
\left(\dfrac{\kk^2}{a(t)^2}+ m^2 \right) |u_k(t)|^2\right] n_B\left(\dfrac{\omega_0(0,k)}{T(0)}\right),\\
p(t)&= \dfrac{1}{\left(2\pi\right)^3}\int\di^3\kk \; \left[  |\dot u_k(t)|^2 - \left(\dfrac{\kk^2}{3 a(t)^2}
+ m^2 \right)|u_k(t)|^2 \right] n_B\left(\dfrac{\omega_0(0,k)}{T(0)}\right),
\end{split}
\end{equation}
where the upper dot denotes a derivative with respect to $t$, and
$$
\omega_0(0,k) = \sqrt{\kk^2+m^2}
$$
according to the eq.~\eqref{omegaxi}.
The differential equations for the mode functions, corresponding to the \eqref{kg2}, read:
\be\label{kg3}
  \ddot{u}_k + 3 H(t) {\dot u}_k + \left(\dfrac{\kk^2}{a^2(t)} + m^2 \right) u_k = 0
\ee
where $H(t) = \dot a(t)/a(t)$ is the expansion rate or Hubble parameter. 

The equation \eqref{kg3} is that of a damped harmonic oscillator, with a friction term proportional
to $H$. This equation is a familiar one for the inflaton field \cite{weinberg}, except that it contains
a term proportional to the covariant momentum squared (see later discussion). The magnitude of the last 
term on the right hand side depends on the mass and the momentum. The relevant magnitude 
of the momentum $\kk$ is selected by the temperature scale $T_0$ in the Bose distribution $n_B$ in equation \eqref{epsilonpt} and by the mass $m$; for instance, if $m \gg T_0$ the regime is non-relativistic and
the relevant momentum scale is $\sqrt{mT_0}$, while if $m \ll T_0$ the regime is ultra-relativistic and
the relevant momentum scale is simply $T_0$. Anyhow, for large scale factors $a(t)$ this contribution 
is generally negligible being suppressed by $1/a^2(t)$, as it is apparent in eq. \eqref{kg3}.

For $H=0$, hence constant $a=1$, the solution of equation \eqref{kg3} is the familiar exponentials
$\exp[\pm \ii \omega_k(0) t]$; after their normalization, the energy density and pressure in equations
\eqref{epsilonpt} boil down to those of an ideal relativistic gas. It is therefore reasonable to suppose 
that for $H$ small compared to the other scale of the problem, that is $m$, the mode functions will be 
similar to the exponentials of the imaginary argument, that is, oscillating functions,and that the 
energy density and pressure will be very close to the classical solutions \eqref{classical} 
obtained with the Boltzmann equation. This case is indeed known as adiabatic expansion 
\cite{mukhanov} and will be further discussed in the next Section.
On the other hand, if the expansion rate is consistently larger than $m, T_0$ for some time, so that 
the typical $k$ selected by the Bose distribution is smaller than $H$, the differential equation is 
that of an overdamped oscillator, with an approximate solution, if $H(t)$
can be approximated as constant for some time, and if the contribution $\kk^2/a(t)^2$ is negligible:
\be\label{expdecay}
  u_k \approx A_k \exp[-m^2 t/3H(t)]
\ee
where $A_k$ is an amplitude. In this case the mode function does not oscillate but it decreases 
exponentially and we have:
$$
|\dot u_k(t)|^2 \approx \dfrac{m^2}{9H^2} m^2 |u_k(t)|^2 \ll m^2 |u_k(t)|^2
$$ 
Therefore, in the equations \eqref{epsilonpt} the kinetic term is much smaller than the
one proportional to the mass, and this makes the pressure negative because the \eqref{epsilonpt}
can be approximated by:
\begin{equation*}
\begin{split}
\varepsilon(t)&\approx \dfrac{1}{\left(2\pi\right)^3}\int\di^3\kk \; m^2 |u_k(t)|^2 
n_B\left(\dfrac{\omega_0(0,k)}{T(0)}\right),\\
p(t)& \approx -  \dfrac{1}{\left(2\pi\right)^3}\int\di^3\kk \; m^2 |u_k(t)|^2
n_B\left(\dfrac{\omega_0(0,k)}{T(0)}\right),
\end{split}
\end{equation*}
implying an equation of state $\varepsilon = - p$. Moreover, since the squared amplitude $|u_k|^2$
is very slowly decaying according to the \eqref{expdecay}, the pressure and the energy density 
are approximately constant over some time during the expansion, thereby mimicking a cosmological 
constant. This is clearly at variance with the classical expressions \eqref{classical} which, 
for a massive particle provide that the energy density decreases as fast as $1/a(t)^3$ and the
pressure as fast as $1/a(t)^5$ for large $a$.

For a massless field, a similar argument applies. In the differential equation \eqref{kg3} with
$m=0$ the form of the solution depends on how $\dot a$ compares to the relevant momentum
scale, that is $T_0$ the temperature at decoupling. If $\dot a \ll T_0$ the friction term can
be neglected, but if $\dot a \gg T_0$ an approximate solution can be obtained again by neglecting
$\ddot{u}_k$ in the \eqref{kg3}:
\be\label{expsol}
  u_k \approx A_k \exp[- \kk^2 t/3a {\dot a}]
\ee
under the assumption that ${\dot a}$ is constant for some time. The above solution makes
the kinetic term in \eqref{epsilonpt} negligible because:
$$
  |\dot u_k|^2 \approx \frac{1}{9} \dfrac{\kk^4}{a^2 \dot a^2} |u_k|^2 \ll 
  \dfrac{\kk^2}{a^2} |u_k|^2
$$
and so, the energy density and pressure \eqref{epsilonpt} can be approximated with:
\begin{equation*}
\begin{split}
\varepsilon(t)&\approx \dfrac{1}{\left(2\pi\right)^3}\int\di^3\kk \; \left(\dfrac{\kk^2}{a(t)^2} \right) 
 |u_k(t)|^2 n_B\left(\dfrac{\omega_0(0,k)}{T(0)}\right),\\
p(t)& \approx \dfrac{1}{\left(2\pi\right)^3}\int\di^3\kk \; \left(- \dfrac{\kk^2}{3 a(t)^2} \right)
|u_k(t)|^2 n_B\left(\dfrac{\omega_0(0,k)}{T(0)}\right),
\end{split}
\end{equation*}
implying an equation of state $p=-(1/3)\varepsilon$. Moreover, since the squared amplitude $|u_k|^2$
is very slowly decaying according to the \eqref{expsol}, the pressure and the energy density 
decrease as fast as $1/a(t)^2$ unlike in the classical expressions \eqref{classical} which, 
for a massless particle, provides that the energy density and pressure both decrease as fast 
as $1/a(t)^4$.

Altogether, the situation can be summarized by saying that the in an expanding universe there
are two opposite limiting regimes for a free field with minimal coupling ($\xi=0$): if the mass 
and the decoupling temperature are much larger than the expansion rate ($H$ for the massive 
case and $\dot a$ for the massless case) then the equation of state is essentially that of a 
classical relativistic system. On the other hand, if the expansion rate is much larger than the
mass and the decoupling temperature, then the pressure becomes as negative as it can be, that is 
$-\varepsilon$ in the massive case and $-\varepsilon/3$ in the massless case. Of course,
between those extremes, there must be a full spectrum of intermediate regimes.

This way of generating a negative pressure is especially interesting because it does not require a 
non-vanishing expectation value of some field, like in the slow-roll inflation model \cite{weinberg}.
In this model, the scalar field operator is required to have a non-vanishing mean value in the quantum 
state of the universe, expressed by some density operator $\wrho$:
$$
 \psi(x) \equiv \Tr (\wrho \, \wpsi(x))
$$
which can be either pure or mixed. By defining the quantum field:
$$
 \delta \wpsi(x) = \wpsi(x) - \psi(x)
$$
it is possible to express the stress-energy tensor operator, in the minimal coupling case, 
in terms of the mean value and the new quantum field $\delta\wpsi(x)$:
\begin{align*}
\wT_{\mu\nu} &= \nabla_\mu \wpsi \nabla_\nu \wpsi - \dfrac{1}{2} g_{\mu\nu} (\nabla \wpsi \cdot \nabla \wpsi
  - m^2 \wpsi^2 ) = \nabla_\mu \psi \nabla_\nu \psi - \dfrac{1}{2} g_{\mu\nu} (\nabla \psi \cdot \nabla \psi
  - m^2 \wpsi^2 ) \\
  & + \nabla_\mu \delta\wpsi \nabla_\nu \delta\wpsi - \dfrac{1}{2} g_{\mu\nu} (\nabla \delta\wpsi 
  \cdot \nabla \delta \wpsi - m^2 \delta\wpsi^2 )
  + \nabla_\mu \delta\wpsi \nabla_\nu \psi + \nabla_\mu \psi \nabla_\nu \delta \wpsi - 
 g_{\mu\nu} (\nabla \delta\wpsi \cdot \nabla \psi - m^2 \psi \, \delta\wpsi  )
\end{align*}
so that by calculating the mean value:
\be\label{meanset}
 \Tr(\wrho \, \wT_{\mu\nu}) = \nabla_\mu \psi \nabla_\nu \psi - \dfrac{1}{2} g_{\mu\nu} (\nabla \psi \cdot \nabla \psi
  - m^2 \psi^2 ) + \Tr \left( \wrho \left[ \nabla_\mu \delta\wpsi \nabla_\nu \delta\wpsi - 
  \dfrac{1}{2} g_{\mu\nu} (\nabla \delta\wpsi   \cdot \nabla \delta \wpsi - m^2 \delta\wpsi^2) \right] \right) \\
\ee
which makes it apparent that the energy density and pressure have a ``classical" contribution from
the expectation value and a fluctuating quantum contribution. The pressure can become negative, in 
this approach, because of the classical term in the \eqref{meanset}, if the field is very slowly 
varying. On the other hand, with the density operator \eqref{densop} the expectation value of the field \eqref{field} is precisely zero and there is no "classical" term in the \eqref{meanset}. The negativity of 
the pressure - as it has been argued - is a possible consequence of the way the quantum field fluctuations
propagate in space and time, i.e. the differential equations of the mode functions \eqref{kg2} as 
well as the definition \eqref{renset}.

In the next Sections, we will demonstrate with concrete examples of solutions of the equations of
motions of the mode functions for given scale functions $a(t)$, that quantum corrections to energy
density and pressure can be large and that pressure can become negative for a macroscopically long 
time, under the conditions of the scale hierarchy mentioned above. However, before doing that, we
discuss in some more detail the so-called adiabatic evolution case, when the Hubble parameter
$H(t)$ is much smaller than the other energy scales.

\section{Adiabatic approximation}
\label{sec:adia}

As has been pointed out in the previous Section, it can be surmised that if the Hubble parameter 
is much smaller than the other energy scales in the problem, that is the mass of the field $m$
and the decoupling temperature $T_0$, the mode functions $v_k$ will be similar to those in flat 
space-time. In this case, in the equation \eqref{kg3} (which applies to the
minimal coupling scenario with $\xi=0$), the friction term proportional to $H(t)$ can be neglected
and one is left with a harmonic oscillator if the scale $a(t)$ is slowly varying. Note that this
approximation may be questionable for very low mass and very long wavelenghts (low $\kk$ values)
or very low mass and very late times where $a(t)$ is large and the $\kk^2/a(t)$ term in the 
equation \eqref{kg3} is negligible. The solutions of the equation \eqref{kg2} can be approximated 
by this very well-known form \cite{weinberg,mukhanov}:
\begin{equation}\label{adia}
    v_k(\eta)\simeq \dfrac{A_k}{\sqrt{2\Omega_k}}\e^{-\ii \int_\eta\Omega_k}+\dfrac{B_k}{\sqrt{2\Omega_k}}
    \e^{\ii\int_\eta\Omega_k},
\end{equation}
where:
\begin{equation*}
    \int_\eta\Omega_k \equiv \int^\eta_{\eta(0)}\di\eta'\Omega_k\left(\eta'\right)
\end{equation*}
The approximation \eqref{adia} is called \emph{adiabatic approximation} for it holds for slowly 
varying scale functions $a(t)$ and, more precisely, it is a good one if:
\begin{equation*}
    \mathcal{A}_k \equiv \dfrac{\Omega'_k}{\Omega^2_k} \ll 1
\end{equation*}
with $\Omega_k$ given in the equation \eqref{Omega}. Thus:
\begin{equation}\label{adiapar}
    \mathcal{A}_k=\dfrac{a'}{a^2}\, \dfrac{m^2-\dfrac{1-6\xi}{2a^2}\left(\dfrac{a'''}{a'}-\dfrac{a''}{a}\right)}
    {\left[m^2+\dfrac{\kk^2}{a^2}-\dfrac{1-6\xi}{a^2}\dfrac{a''}{a}\right]^{3/2}}
    = H \, \dfrac{m^2-\dfrac{1-6\xi}{2a^2}\left(\dfrac{a'''}{a'}-\dfrac{a''}{a}\right)}
    {\left[m^2+\dfrac{\kk^2}{a^2}-\dfrac{1-6\xi}{a^2}\dfrac{a''}{a}\right]^{3/2}}
\end{equation}
where we used the equality $H = a'/a^2$ for the Hubble parameter in the conformal coordinates. 
For most expansion dynamics in the limit $a \to \infty$ in the massive case one has:
$$
 \mathcal{A}_k \simeq \dfrac{H}{m}
$$
which makes it apparent that the adiabatic approximation applies when the Hubble parameter
is much smaller than the mass. On the other hand, for a massless field, the adiabaticity 
parameter is more sensitive to the expansion dynamics, so we will confine ourselves at this time
to the massive case; the massless case will be dealt with separately.

The two coefficients $A_k$ and $B_k$ in the equation \eqref{adia} can be obtained by matching
the actual solution of the equation of motion \eqref{kg2} to the form \eqref{adia} when the 
adiabatic approximation becomes viable. Note that the Wronskian of the form \eqref{adia} reads:
$$
 v_kv'^*_k-v'_kv^*_k=i\left(\left|A_k\right|^2-\left|B_k\right|^2\right)
$$
implying that the coefficients fulfill the identity:
$$
 |A_k|^2 - |B_k|^2 = 1
$$
The magnitude of $B_k$, hence of $A_k$, depends on the initial conditions, i.e. at the decoupling
time. If the adiabatic approximation holds from the decoupling, i.e. with ${\mathcal A}_k(0) \ll 1$, 
the evolution of the field is always adiabatic and the component proportional to $B_k$, with the 
initial conditions in eq.~\eqref{condin} turns out to be proportional to $\mathcal{A}_k$ 
(see Appendix \ref{app:adiadec} for the derivation):
\be\label{Badia}
\begin{split}
 A_k& \simeq 1-\dfrac{i\left(3-12\xi\right)}{4}{\mathcal A}_k(0) \\
 B_k & \simeq -\dfrac{\ii\left(3-12\xi\right)}{4} \, \mathcal{A}_k(0)
 \end{split}
\ee
with:
\be\label{adzero}
 \mathcal{A}_k(0) =  H(0) \, \dfrac{m^2-\dfrac{1-6\xi}{2} \left(\dfrac{a'''(0)}{a'(0)}-a''(0)\right)}
    {\left[m^2+\kk^2-(1-6\xi)a''(0)\right]^{3/2}} \lesssim 1
\ee
Conversely, if $\mathcal{A}_k(0) \gtrsim 1$,it is reasonable to expect a larger coefficient $B_k$ 
hence a sizeable mixing between the positive and negative frequency components.

In order to calculate energy density and pressure in the adiabatic approximation, we need the quadratic 
forms of the mode functions, which can be readily obtained from the \eqref{adia}:
\begin{equation*}
    \begin{split}
\left|v_k\right|^2&=\dfrac{1}{2\Omega_k}\left(\left|A_k\right|^2+\left|B_k\right|^2\right)+
\dfrac{1}{\Omega_k}\mbox{Re}\left[A_kB^*_k\e^{-2i\int_\eta\Omega_k}\right],\\
\left|v'_k\right|^2&=\dfrac{1}{2}\left(\Omega_k+\dfrac{\Omega'^2_k}{4\Omega^3_k}\right)
\left(\left|A_k\right|^2+\left|B_k\right|^2\right)+\left(-\Omega_k+\dfrac{\Omega'^2_k}{4\Omega^3_k}\right)
\mbox{Re}\left[A_kB^*_k\e^{-2i\int_\eta\Omega_k}\right]+\left(\dfrac{\Omega'_k}{\Omega_k}\right)
\mbox{Im}\left[A_kB^*_k\e^{-2i\int_\eta\Omega_k}\right],\\
 v'_kv^*_k&=\left(-\dfrac{\Omega'_k}{4\Omega^2_k}\right)\left(\left|A_k\right|^2+\left|B_k\right|^2\right)
 +\dfrac{i}{2}\left(\left|A_k\right|^2-\left|B_k\right|^2\right)-\mbox{Im}\left[A_kB^*_k\e^{-2i\int_\eta\Omega_k}\right]
 -\left(\dfrac{\Omega'_k}{2\Omega^2_k}\right)\mbox{Re}\left[A_kB^*_k\e^{-2i\int_\eta\Omega_k}\right].
    \end{split}
\end{equation*}
By using the above expressions, one can obtain the function $\omega_k K_k(\eta)$ in the eq. 
\eqref{funzKGAMMA}:
\be\label{omegaK}
\begin{split}
 \omega_kK_k(\eta) 
 &= \Omega_k \left( \left|A_k\right|^2+\left|B_k\right|^2 \right) + \dfrac{1-6\xi}{2\Omega_k}
 \left[ \dfrac{a'}{a} \Omega_k\mbox{Im}\left[ A_k B^*_k \e^{-2i\int_\eta\Omega_k} \right] \right. \\
 &+ \left. \left( \dfrac{a'^2}{a^2}+\dfrac{a''}{a}\right) \left( |A_k|^2+|B_k|^2+2\mbox{Re}
 \left[A_kB^*_k \e^{-2\ii\int_\eta\Omega_k} \right] \right) \right] + {\cal O}({\cal A}_k)
\end{split}
\ee
Similarly, for the function $\Gamma_k$:
\be\label{omegaGamma}
    \begin{split}
    \omega_k\Gamma_k&=\left[\dfrac{\kk^2}{3\Omega_k}+\dfrac{\left(1-6\xi\right)}{2\Omega_k}
    \left(\dfrac{a'^2}{a^2}-\dfrac{a''}{a}\right)\right]\left(|A_k|^2+|B_k|^2\right)\\
    &+\left[\left(8\xi-1\right)\Omega_k+\dfrac{1}{\Omega_k}\left(-\dfrac{\kk^2}{3}-m^2a^2+\left(1-6\xi\right)\dfrac{a'^2}{a^2}\right)\right]\mbox{Re}\left[A_kB^*_k\e^{-2i\int_\eta\Omega_k}\right]\\
    &+\dfrac{\left(1-6\xi\right)a'}{a}\mbox{Im}\left[A_kB^*_k\e^{-2i\int_\eta\Omega_k}\right]+O\left(\mathcal{A}_k\right).
\end{split}
\ee
These expressions can be further approximated by taking into account the adiabaticity condition 
$H \ll m$. In this regime we have:
\be\label{omegaapprox}
   \Omega_k^2 = \kk^2 + m^2a^2-\left(1-6\xi\right)\dfrac{a''}{a} \simeq \kk^2 + m^2a^2
\ee
Indeed, since:
$$
 \dfrac{a''}{a} = \ddot{a} a + {\dot a}^2 = \ddot{a} a + H^2 a^2
$$
the term proportional to $(1-6\xi)$ in the \eqref{omegaapprox} can be neglected provided that 
$m^2 \gg \ddot a/a$ which applies even in the case of an exponential expansion. Besides:
$$
  \omega_k^2 \simeq \kk^2 + m^2a^2
$$
By using similar arguments it can be shown that the term proportional to $(1-6\xi)$ on the 
right hand side of the equation \eqref{omegaK} can be neglected, hence:
\be\label{omegaKdom}
 \omega_kK_k \simeq \omega_k \; (1+2 |B_k|^2)
\ee
where we have used the identity $|A_k| - |B_k|^2 = 1$. One can recognize in the first term of the right 
hand side of \eqref{omegaKdom} the classical expressions of the integrand of the energy density in 
eq.~\eqref{classical}. 
Also, the first corrections to the classical expression is an enhancement factor proportional to 
$2|B_k|^2$ which can be interpreted as a spontaneous particle production \cite{Kolb:2023ydq}. Altogether,
in the adiabatic regime:
$$
a^4\varepsilon\simeq \dfrac{1}{(2\pi)^3}\int \di^3\kk \; 
\omega_k\left(1+2\left|B_k\right|^2\right)n_B\left(\dfrac{\omega_0\left(0,\kk\right)}{T_0}\right),
$$
Similarly, by using again the condition $H \ll m$, the right hand side of equation 
\eqref{omegaGamma} can be approximated by:
\be\label{omegaGammadom}
\begin{split}
 \omega_k\Gamma_k &\simeq\dfrac{\kk^2}{3\omega_k}\left(1 +2|B_k|^2 \right) + 
 \dfrac{\left(1-6\xi\right)}{2\omega_k} (\dot H a^2 + H^2 a^2) \left(1 +2|B_k|^2 \right) \\
&-\dfrac{\left(1-8\xi\right)\omega^2_k+m^2a^2+\kk^2/3}{\omega_k}\mbox{Re}
\left[A_kB^*_k \e^{-2\ii\int_\eta\Omega_k}\right] + a H \left(1-6\xi\right)
\mbox{Im}\left[A_kB^*_k\e^{-2\ii\int_\eta\Omega_k}\right]
\end{split}
\ee
The classical term $\kk^2/3 \Omega_k$ enhanced by a factor $(1 + 2 |B_k|^2)$ shows up in the 
equation \eqref{omegaGammadom} as well. There is, however, a second term proportional to $(1 + 2 |B_k|^2)$
which depends on the dynamics of the expansion in the long run, i.e. how $H^2 a^2 + \dot H a^2$ 
compares to the typical momentum scale, which is driven by both mass $m$ and decoupling temperature $T_0$. 
However, unlike the energy density, the pressure integrand receives oscillating corrections in 
\eqref{omegaGammadom}, which vanish only if $B_k=0$. In the long run, the frequency of the 
oscillating terms is fixed by the mass of the field. In fact, by using the eq.~\eqref{omegaapprox} 
and the relation between cosmological and conformal time $\di \eta = \di t/a$ one readily obtains:
\begin{equation*}
    -\ii \int\di\eta'\Omega_k(\eta') \simeq -\ii \int\dfrac{\di t}{a}ma=-\ii m t,
\end{equation*}
On the other hand, the amplitude of the oscillating terms requires the calculation of the coefficients 
$A_k$ and $B_k$, whose magnitude depends, as has been discussed, on the conditions prevailing at the 
decoupling. Anyhow, in the long time limit where $a \to \infty$ the $\omega_k$ gets dominated by the 
mass term $m a(t)$ and the \eqref{omegaKdom},\eqref{omegaGammadom} can be approximated by:
\begin{equation*}
    \begin{split}
        \omega_kK_k&\sim m a\left(1+2\left|B_k\right|^2\right),\\
        \omega_k\Gamma_k&\sim-2\left(1-4\xi\right)m a \, \mbox{Re}\left(A_kB^*_k \e^{-2imt}\right),
    \end{split}
\end{equation*}
Such expressions imply that in the adiabatic approximation, in the long time limit, the ratio between 
pressure and energy density is oscillating with a frequency $m/\pi$ according to:
\begin{equation}\label{woscill}
    \lim_{t\to\infty}\dfrac{p}{\varepsilon}=-2\left(1-4\xi\right)\dfrac{\int\di^3\kk \, \mbox{Re}
        \left( A_k B^*_k \e^{-2\ii m t} \right) n_B\left(\omega_\xi\left(0,T_0\right)\right)}
        {\int\di^3\kk \, \left(1+2\left|B_k\right|^2\right)n_B\left(\omega_\xi\left(0,T_0\right)\right)}.
\end{equation}
This oscillation is but a sheer quantum effect and it is macroscopically unobservable unless the
mass is extraordinary small. If the adiabatic approximation applies to decoupling as well, it
is possible to use the estimates \eqref{Badia} for the mixing coefficients and the equation 
\eqref{woscill} becomes:
\begin{equation*}
    \lim_{t\to\infty}\dfrac{p}{\varepsilon} \simeq -\frac{3}{2}\left(1-4\xi\right)^2
        \left[ \dfrac{\int\di^3\kk \, \mathcal{A}_k(0) n_B \left(\omega_\xi\left(0,T_0\right)\right)}
        {\int\di^3\kk \, n_B\left(\omega_\xi\left(0,T_0\right)\right)} \right] \, \sin (2mt).
\end{equation*}
If all the higher order derivative terms appearing in \eqref{adzero} are also negligible 
compared to the mass, we can approximate the adiabatic parameter at the decoupling as
(see Appendix B):
$$
  \mathcal{A}_k(0) \simeq H(0) \frac{m^2}{\epsilon^3_k} 
$$
with $\epsilon_k = \sqrt{\kk^2+m^2}$; hence, if the decoupling occurs in a non-relativistic regime
($T_0 \ll m$) one has the simple law:
\begin{equation*}
    \lim_{t\to\infty}\dfrac{p}{\varepsilon} \simeq -\frac{3}{2}\left(1-4\xi\right)^2 
    \frac{H(0)}{m} \, \sin (2mt).
\end{equation*}
%

\section{De Sitter universe}
\label{sec:desitter}

The most interesting feature of the De Sitter universe is that its expansion rate allows an 
evolution of the field that is asymptotically non-adiabatic and for which it is therefore not 
possible to apply the approximation \eqref{adia}.

A cosmological de Sitter universe is a FLRW metric with scale factor which is exponential in cosmological 
time:
\begin{equation*}
    a(t)=\e^{H\left(t-t_0\right)}.
\end{equation*}
Such spacetime has a constant Hubble parameter $H=\dot{a}/a$, and an accelerated expansion $\ddot{a}>0$. 
The conformal time reads:
\begin{equation}\label{etat}
    \eta(t)=\dfrac{1}{H}\left(1-\e^{-H\left(t-t_0\right)}\right),
\end{equation}
and converges in the far future:
\begin{equation*}
    \lim_{t\to\infty}\eta(t)=\dfrac{1}{H}.
\end{equation*}
The convergence of $\eta(t)$ signals the presence of a cosmological horizon whose asymptotic comoving 
length is $1/H$. The scale factor in the conformal time is:
\be\label{adesitter}
    a(\eta)=\dfrac{1}{1-H\eta}.
\ee
then:
\be\label{aderdesitter}
    a'=\dfrac{H}{\left(1-H\eta\right)^2} \qquad  a''=\dfrac{2H^2}{\left(1-H\eta\right)^3}
    \qquad a'''=\dfrac{6H^3}{\left(1-H\eta\right)^4}.
\ee
We will now study the solutions of the Klein-Gordon field equations and the resulting energy
density and pressure in the massive and massless case separately.

\subsection{Massive case}

For a de Sitter universe the ratio $H/m$ is constant throughout the evolution. By using the
equation \eqref{adiapar} and the \eqref{adesitter}, \eqref{aderdesitter} it can be readily shown
that:
\begin{equation}\label{adiapards}
    \mathcal{A}_k(\eta) =\frac{H}{m}\frac{1-2\left(1-6\xi\right)\frac{H^2}{m^2}}
    {\left[1-2\left(1-6\xi\right)\frac{H^2}{m^2}+k^2\left(1-H\eta\right)^2\right]^{3/2}}.
\end{equation}
The time evolution of the adiabatic parameter depends on the sign of $1-2\left(1-6\xi\right)H^2/m^2$. 
If it is positive, then $\mathcal{A}_k$ is a monotonically increasing function and its asymptotic 
value is given by:
\begin{equation*}
    \lim_{\eta\to 1/H}\mathcal{A}_k=\frac{H}{m}\frac{1}{\sqrt{1-2\left(1-6\xi\right)H^2/m^2}}.
\end{equation*}
On the other hand, if $1-2\left(1-6\xi\right)H^2/m^2<0$, we may have imaginary values of the adiabatic 
parameter at late times, signifying a non-adiabatic expansion. Hence, if the following conditions 
are met:
\begin{align*}
  &  1-2\left(1-6\xi\right)H^2/m^2 \ge 0 \implies \frac{H}{m} \le \sqrt{\frac{1}{2(1-6\xi)}} \\
  &  \frac{H}{m}\frac{1}{\sqrt{1-2\left(1-6\xi\right)H^2/m^2}} \le 1 \implies 
  \frac{H}{m} \le \sqrt{\frac{1}{3(1-4\xi)}}
\end{align*}
we have $\mathcal{A}_k(0) < \mathcal{A}_k(\infty) \le 1$ which corresponds to an adiabatic expansion
throughout. Regardless of the value of $\xi$ within the interval $[0,1/6]$, if $H \ll m$ the regime will 
always be adiabatic.
In conclusion, for a massive field in de Sitter space-time, there are two possible regimes: 
\begin{equation}    
    \qquad \dfrac{H}{m}\ll1\implies\ \mbox{Adiabatic\ regime throughout} \qquad\qquad
    \dfrac{H}{m}\gtrsim1\implies\ \mbox{Non-adiabatic\ regime};
\end{equation}   

For a massive field in a de Sitter universe the free Klein-Gordon equation in conformal time
reads:
\begin{equation}\label{DeSitterKG}
    \dfrac{\di v_k(y)}{\di y^2}+\left[\dfrac{\kk^2}{H^2}+\dfrac{m^2/H^2-2\left(1-6\xi\right)}{y^2}\right]v_k(y)=0,
\end{equation}
with
\be\label{ydef}
 y \equiv 1-H\eta = \frac{1}{a(\eta)},\qquad y(t_0)=1,\quad y(\infty)=0.
\ee
It is well known that the general solution is a linear combination of Hankel functions 
$\huno(z)$ and $\hdue(z)$ where $z=\kk y/H$ and the order $\nu$ reads:
\begin{equation}\label{orderDeSitter}
    \nu=\dfrac{1}{2}\sqrt{9-48\xi-\dfrac{4m^2}{H^2}},
\end{equation}
The order can be a real number or a pure imaginary number depending on the magnitude of the term 
$-48\xi-4m^2/H^2$ and it is independent from the scale of the mode $\kk$. The effective Hamiltonian of 
the field turns out to be bounded from below at any time if $\xi\in\left[0,1/6\right]$,  \cite{becarose1}
thus the order $\nu$ is real when
\begin{equation*}
    \dfrac{H}{m} \ge \dfrac{3}{2}\sqrt{1-\dfrac{16\xi}{3}} \implies \dfrac{H}{m} \ge \dfrac{3}{2} \;\; {\rm for}\; {\xi=0}
    \qquad \dfrac{H}{m}\ge\dfrac{1}{2} \;\; {\rm for} \; {\xi=1/6}
\end{equation*}
Hence, in the adiabatic regime with $H/m\ll1$ the order is always a pure imaginary number while for the 
non-adiabatic case $H\gtrsim m$ the order is real. 

\subsubsection{Adiabatic case}

We first consider the case of an adiabatic regime. From (\ref{orderDeSitter}) the order is a pure 
imaginary number $\nu=i\mu$:
\begin{equation*}
    \mu\equiv\dfrac{1}{2}\sqrt{\dfrac{4m^2}{H^2}+48\xi-9}\simeq \dfrac{m}{H}\gg1.
\end{equation*}
Using known relations (see Appendix A, eq.~\eqref{HunoHdueIM}) for Hankel functions of pure imaginary 
order we obtain the normalized solution of the equation (\ref{DeSitterKG}):
\begin{equation}
    v_k(y)=C_k\sqrt{\dfrac{\pi}{4H}}\, \e^{-\mu\pi/2}\sqrt{y}\, \hunoC\left(\dfrac{\kk y}{H}\right)+
    D_k\sqrt{\dfrac{\pi}{4H}}\, \e^{\mu\pi/2}\sqrt{y}\, \hdueC\left(\dfrac{\kk y}{H}\right),
\end{equation}
where $C_k$, and $D_k$ are fixed by the initial conditions (\ref{condin}): 
\begin{equation}\label{CoeffOrdineIm}
\begin{split}
C_k&=\sqrt{\dfrac{2H}{\pi\omega_\xi(0)}}\dfrac{\e^{\mu\pi/2}}{\hunoC\left(\kk/H\right)}
  +\sqrt{\dfrac{\pi\omega_\xi(0)}{8H}} \e^{\mu\pi/2}\hdueC\left(\kk/H\right)
  \left[1+\ii \frac{H}{2\omega_\xi\left(0\right)}\left(3-12\xi+\dfrac{2\kk}{H\hunoC\left(\kk/H\right)}
  \dfrac{\di \hunoC\left(\kk/H\right)}{\di\left(\kk/H\right)}\right)\right],\\
D_k&=-\sqrt{\dfrac{\pi\omega_\xi(0)}{8H}} \e^{-\mu\pi/2}\hunoC\left(\kk/H\right)\left[1+\ii \frac{H}{2\omega_\xi\left(0\right)}\left(3-12\xi+\dfrac{2\kk}{H\hunoC\left(\kk/H\right)}
  \dfrac{\di \hunoC\left(\kk/H\right)}{\di\left(\kk/H\right)}\right)\right].
\end{split}
\end{equation}
with $\left|C_k\right|^2-\left|D_k\right|^2=1$ as required by the Wronskian condition and $\omega_\xi(\eta)$ given by \eqref{omegaxi}.
In the late time limit, corresponding to $y \to 0$, the Hankel function can be replaced by 
an asymptotic expansion for large imaginary orders and for small argument (see eq.~\eqref{HnuIsmall} 
and the relation \eqref{HunoHdueIM} in the Appendix A):
\begin{equation*}
\begin{split}
v_k(y)&\sim \dfrac{C_k}{\sqrt{2m}}\sqrt{y}\left\{\exp\left[\ii\mu\left(\log\left(\dfrac{\kk y}{2H}\right)
-\log\left(\dfrac{\mu}{e}\right)\right)-\dfrac{\ii \pi}{4}\right]\right\}\\
&+\dfrac{D_k}{\sqrt{2m}}\sqrt{y}\left\{\exp\left[-\ii \mu\left(\log\left(\dfrac{\kk y}{2H}\right)+
\log\left(\dfrac{\mu}{e}\right)\right)+\dfrac{\ii\pi}{4}\right]\right\}.
\end{split}
\end{equation*}
This solution is essentially the adiabatic approximation (\ref{adia}), as expected. To demonstrate
it, one should take into account that for $\mu\simeq m/H$ the logarithms in the exponent can 
be written in a compact form as $\log\left(e\kk y/2m\right)$, which implies the following 
expansion for the mode:
\begin{equation}\label{asymptv}
    v_k(y)\sim \dfrac{C_k}{\sqrt{2m}}\sqrt{y}\, \e^{\ii\mu\log\left(e\kk y/2m \right)-
    \ii\pi/4}+ \dfrac{D_k}{\sqrt{2m}}\sqrt{y} \, \e^{-\ii \mu\log\left( e\kk y/2m\right)+
    \ii\pi/4}.
\end{equation}
The exponent of the adiabatic solution can be written, by changing the integration variable
from $\eta$ to $y$:
$$
\int^\eta_{0}\di\eta'\Omega_k\left(\eta'\right)=-H^{-1}\int^y_{1}\di y'\Omega_k\left(y'\right)
$$
whence:
\begin{equation*}
    \dfrac{\exp\left[-\ii \int_\eta\Omega_k\right]}{\sqrt{2\Omega_k}}
    =\dfrac{\sqrt{y} \, \exp\left[i\mu\int^y_1\di y'\sqrt{\dfrac{\kk^2}{m^2}+
    \dfrac{1-2\left(1-6\xi\right)\mu^{-2}}{y'^2}}\right]}{\sqrt{2\sqrt{\kk^2y^2+
    m^2\left[1-2\left(1-6\xi\right)\mu^{-2}\right]}}},
\end{equation*}
which for late times $y\to0$ and $\mu\gg1$ can be well approximated by:
\begin{equation*}
    \dfrac{\e^{-\ii \int_\eta\Omega_k}}{\sqrt{2\Omega_k}}\simeq \dfrac{\sqrt{y}}{\sqrt{2m}}
    \e^{\ii\mu\log\left(e\kk y/2m\right)}\e^{-\ii \mu\log\left(e\kk/2m\right)}.
\end{equation*}
Therefore, we can rewrite the asymptotic form of the modes in eq.~\eqref{asymptv}, as:
\begin{equation*}
    v_k(y)\sim \dfrac{C_k}{\sqrt{2\Omega_k}}\e^{\ii \left(\mu\log\left(e\kk/2m\right)-\pi/4\right)}
    \e^{-\ii \int_\eta\Omega_k}+\dfrac{D_k}{\sqrt{2\Omega_k}}\e^{-\ii \left(\mu\log\left(e\kk/2m\right)
    -\pi/4\right)}\e^{\ii\int_\eta\Omega_k}.
\end{equation*}
whence the coefficients $A_k$ and $B_k$ are readily obtained
\begin{equation}\label{aandb}
\begin{split}
    A_k &= C_k \, \e^{\ii\mu\log\left(e\kk/2m\right)-\ii \pi/4}\\
    B_k &= D_k \, \e^{-\ii\mu\log\left(e\kk/2m\right)+\ii \pi/4},
\end{split}
\end{equation}

It is also possible to find asymptotic expressions of the coefficients $C_k$ and $D_k$ in the equation
\eqref{CoeffOrdineIm}. The argument of the coefficient can be rewritten as:
$$
 \dfrac{\kk}{H} = \dfrac{m}{H} \, \dfrac{\kk}{m} \simeq \mu \dfrac{\kk}{m}
$$
For a finite $u \equiv \kk/m$, the argument of the Hankel functions in \eqref{CoeffOrdineIm}
is thus similar to the order, hence one needs the asymptotic expansion of Hankel functions 
for large order and large arguments (see equation \eqref{HnuIhigh} in the Appendix A):
\begin{equation}\label{candd}
\begin{split}
C_k& \simeq \left[1+\dfrac{\ii}{4\mu} \left( \dfrac{3-12\xi}{\sqrt{1+u^2}}-\dfrac{u}{\left(1+u^2\right)^{3/2}} 
\right) \right] \e^{\ii [ \pi/4-\mu\zeta(u)]},\\
D_k& \simeq -\dfrac{\ii}{4\mu}\left(\dfrac{3-12\xi}{\sqrt{1+u^2}}-\dfrac{u}{\left(1+u^2\right)^{3/2}}\right) 
\e^{-\ii\left[\pi/4-\mu\zeta(u) \right]}
\end{split}
\end{equation}
with $\zeta(u)$ given by \eqref{zeta(x)}
and, from \eqref{aandb}
\begin{equation}\label{aandb2}
\begin{split}
    A_k &\sim \left[1+\dfrac{\ii}{4\mu}\left(\dfrac{3-12\xi}{\sqrt{1+u^2}}-
    \dfrac{u}{\left(1+u^2\right)^{3/2}}\right)\right]\e^{\ii\mu \left(\log(\e u/2)-\zeta(u)\right)}\\
    B_k &\simeq-\dfrac{\ii}{4\mu}\left(\dfrac{3-12\xi}{\sqrt{1+u^2}}-\dfrac{u}{\left(1+u^2\right)^{3/2}}
    \right)\e^{-\ii\mu \left(\log(\e u/2)-\zeta(u)\right)}
\end{split}
\end{equation}
From the equations \eqref{candd} and \eqref{aandb2} it can be concluded that the coefficients 
have a leading behaviour as a function of $\mu \simeq m/H$ is:
\begin{equation*}
    \left|B_k\right|^2 = \left|D_k\right|^2 \propto\mu^{-2} 
    \qquad A_k B^*_k \propto C_k D^*_k \propto \mu^{-1},
\end{equation*}
and so the quantum corrections in the energy density and pressure are very small, as expected for an
adiabatic decoupling.

\begin{figure}[H]
\centering
\begin{subfigure}{.45\textwidth}
  \centering
  \includegraphics[width=.99\linewidth]{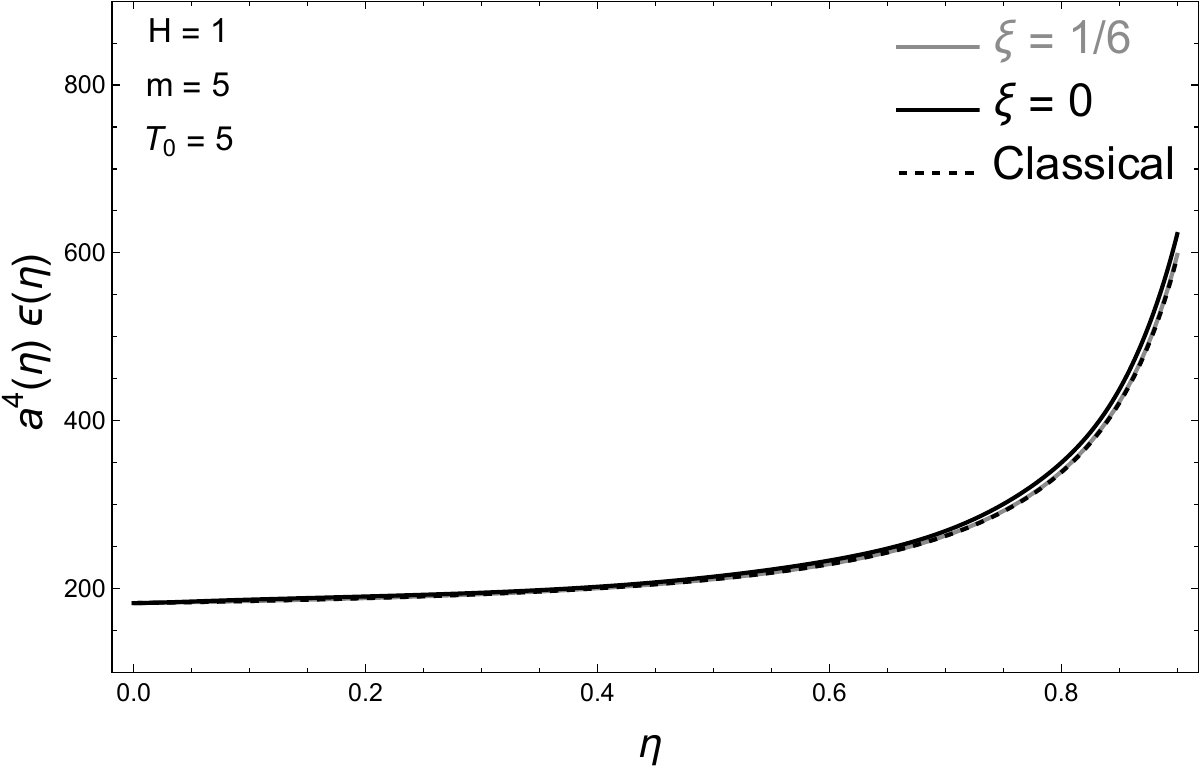}
  \caption{Energy density}
  \label{dsad1}
\end{subfigure}%
\hspace{1cm}
\begin{subfigure}{.45\textwidth}
  \centering
  \includegraphics[width=.99\linewidth]{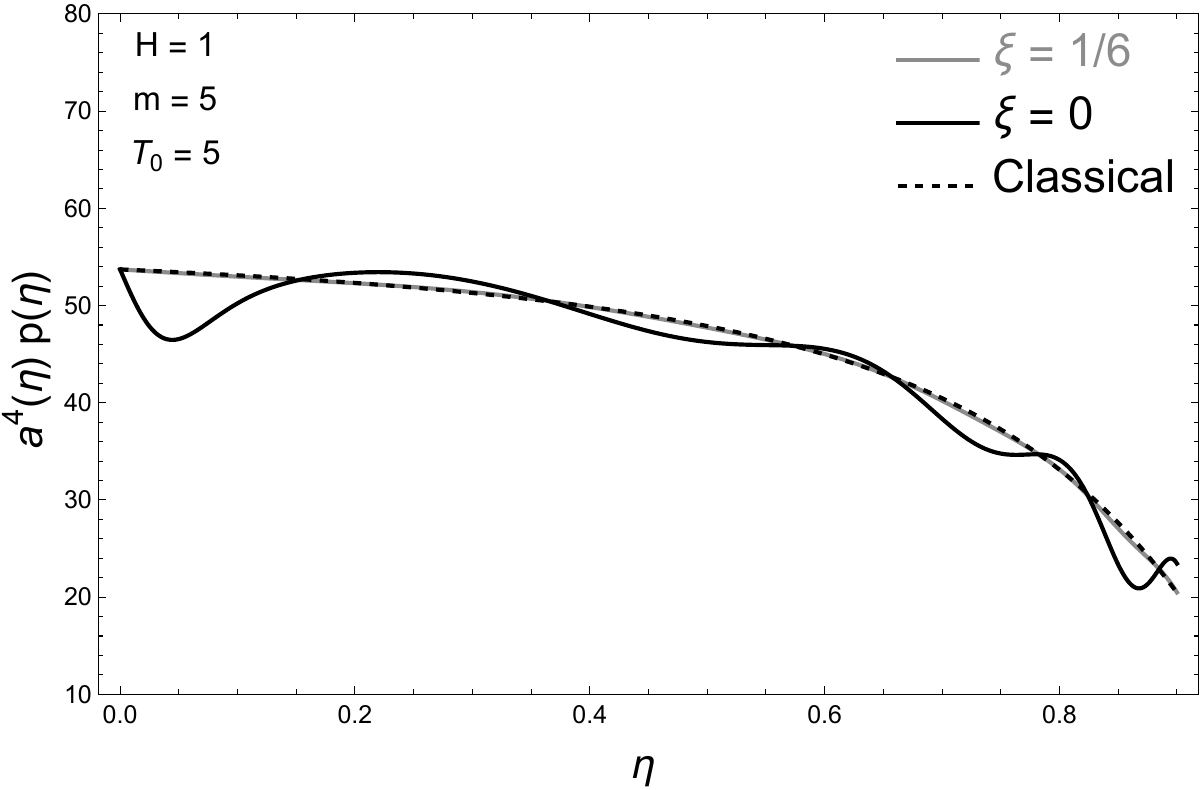}
  \caption{Pressure}
  \label{dsad2}
\end{subfigure}
\caption{Energy density and pressure as a function of conformal time for a de Sitter universe
for $H=1$, $m=5$ and $T_0=5$ (arbitrary units).}
\end{figure}
\begin{figure}[H]
\centering
\includegraphics[width=.55\textwidth]{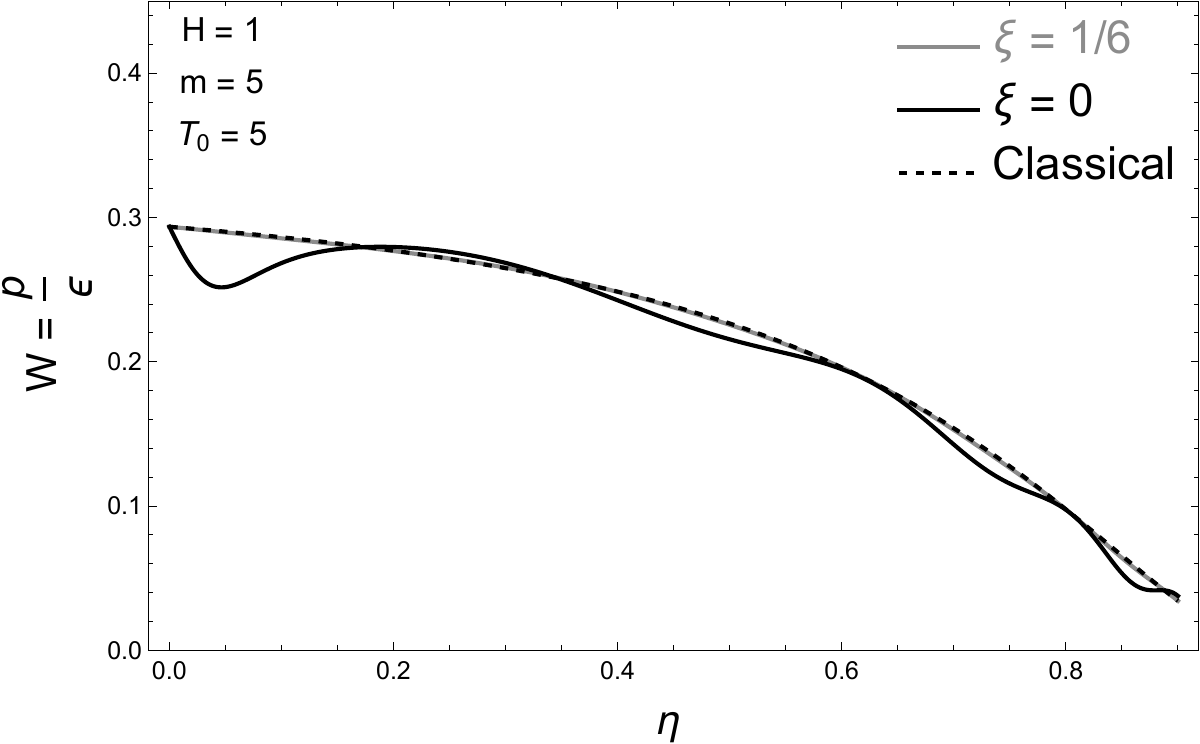}
\caption{Ratio of pressure and energy density as a function of conformal time for a 
de Sitter universe for $H=1$, $m=5$ and $T_0=5$ (arbitrary units).}
\label{dsad3}
\end{figure}

In figures~\ref{dsad1},\ref{dsad2} and \ref{dsad3} the energy density, pressure and their ratio
calculated from the exact expressions of the mode functions are shown as a function of conformal 
time for a ratio $m/H = 5$ and $T_0 = m$. Indeed, it can be seen that the quantum corrections 
to the classical expressions are small and oscillating around them, as predicted by the adiabatic 
approximation.

\subsubsection{Non-adiabatic case}

In the case of $H/m \gtrsim 1$, the order $\nu$ is real, as we have seen and the solution of 
the \eqref{DeSitterKG}
reads:
\begin{equation}\label{SolDeSittMassNonAd}
    v_k(y) =C_k \sqrt{\dfrac{\pi}{4H}}\sqrt{y}\, \huno\left(\dfrac{\kk y}{H}\right)+
    D_k \sqrt{\dfrac{\pi}{4H}}\sqrt{y} \, \hdue\left(\dfrac{\kk y}{H}\right).
\end{equation}
Again, the coefficients $C_k$ and $D_k$ are fixed by the initial conditions \eqref{condin}:
\begin{equation}\label{CoefficientiDeSittMassNonAd}
\begin{split}
C_k&=\sqrt{\dfrac{2H}{\pi\omega_\xi(0)}}\dfrac{1}{\huno\left(\kk/H\right)}+\sqrt{\dfrac{\pi\omega_\xi(0)}{8H}}\hdue\left(\kk/H\right)\left[1+\dfrac{iH}{2\omega_\xi(0)}\left(3-12\xi+2\nu-2\dfrac{\kk}{H}
\dfrac{\hunopiu\left(\kk/H\right)}{\huno\left(\kk/H\right)}\right)\right],\\
D_k&=-\sqrt{\dfrac{\pi\omega_\xi(0)}{8H}}\huno\left(\kk/H\right)\left[1+\dfrac{iH}{2\omega_\xi(0)}\left(3-12\xi+2\nu-2\dfrac{\kk}{H}
\dfrac{\hunopiu\left(\kk/H\right)}{\huno\left(\kk/H\right)}\right)\right].
\end{split}
\end{equation}
As required by the Wronskian condition $\left|C_k\right|^2-\left|D_k\right|^2=1$.

For late times $y\to0$ and one can use the expansions of the Hankel functions for 
small arguments and fixed real order (see Appendix A, eq.~\eqref{HnuRsmall}):
\begin{equation}\label{vkapproxds}
    v_k(y)\sim-\sqrt{\dfrac{\pi}{4H}}\dfrac{2^\nu \ii \Gamma(\nu)E_k}{\pi}
    \left(\dfrac{H}{\kk}\right)^\nu\dfrac{1}{y^{\nu-1/2}},
\end{equation}
where the coefficient $E_k$ is defined as:
\begin{equation*}
\begin{split}
E_k \equiv \left(C_k-D_k\right)= 
\sqrt{\dfrac{4H}{\pi}}&\left\{\dfrac{1}{\sqrt{2\omega_\xi(0)}\huno\left(\kk/H\right)}
+\dfrac{\pi\sqrt{2\omega_\xi(0)}\left(\huno\left(\kk/H\right)+\hdue\left(\kk/H\right)\right)}
{8 H}\times\right.\\
    &\left.\times\left[1+\ii\dfrac{H}{2\omega_\xi(0)}\left(3-12\xi+2\nu-2\dfrac{\kk}{H}
    \dfrac{\hunopiu\left(\kk/H\right)}{\huno\left(\kk/H\right)}\right)
    \right]\right\}.
\end{split}
\end{equation*}
Indeed, the \eqref{vkapproxds} becomes a good approximation when the argument of the Hankel 
functions in \eqref{SolDeSittMassNonAd} $\kk y/H \ll 1$, that is when, by using the definitions
of $y$ and the equation \eqref{etat}, 
$$
    t-t_0 \gg \dfrac{1}{H} \log\dfrac{\kk}{H}
$$
The time scale depends, besides the Hubble parameter, on the logarithm of the typical momentum 
which is related to the decoupling temperature and the mass.

The approximate form of the quadratic combinations can be readily found on the basis
of \eqref{vkapproxds}:
\begin{equation*}
    \begin{split}
\left|v_k\right|^2&\sim\dfrac{2^{2\nu-2}\Gamma^2(\nu)\left|E_k\right|^2}
 {H\pi}\left(\dfrac{H}{\kk}\right)^{2\nu}\dfrac{1}{y^{2\nu-1}},\\
 \left|\dfrac{\di v_k}{\di y}\right|^2&\sim\dfrac{2^{2\nu-2}\Gamma^2(\nu)
\left(\dfrac{1}{2}-\nu\right)^2\left|E_k\right|^2}{H\pi}\left(\dfrac{H}{\kk}\right)^{2\nu}
\dfrac{1}{y^{2\nu+1}},\\
\dfrac{\di v_k}{\di y}v^*_k&\sim\dfrac{2^{2\nu-2}\Gamma^2(\nu)
\left(-\dfrac{1}{2}+\nu\right)\left|E_k\right|^2}{H\pi}\left(\dfrac{H}{\kk}\right)^{2\nu}
\dfrac{1}{y^{2\nu}}.
    \end{split}
\end{equation*}
Plugging the above expressions in the equation \eqref{funzKGAMMA}, one obtains the 
asymptotic form for the integrands of $\varepsilon$ and $p$:
\begin{equation}\label{AsymptoticIntegrandNonAdDS}
    \begin{split}
\omega_kK_k&\simeq\dfrac{2^{2\nu-2}\Gamma^2(\nu)\left|E_k\right|^2}{H\pi}
\left(\dfrac{H}{\kk}\right)^{2\nu}\left[\dfrac{m^2}{H^2}+\left(\dfrac{1}{2}-\nu\right)^2
+2\left(1-\nu\right)\left(1-6\xi\right)\right]\dfrac{H^2}{y^{2\nu+1}},\\
\omega_k\Gamma_k&\simeq\dfrac{2^{2\nu-2}\Gamma^2(\nu)\left|E_k\right|^2}{H\pi}
\left(\dfrac{H}{\kk}\right)^{2\nu}\left\{\left(1-4\xi\right)\left[-\dfrac{m^2}{H^2}
+\left(\dfrac{1}{2}-\nu\right)^2+2\left(1-6\xi\right)\right]-2\nu\left(1-6\xi\right)\right\}
\dfrac{H^2}{y^{2\nu+1}}.
    \end{split}
\end{equation}
Since $y=1/a$ according to the \eqref{ydef}, it turns out from the equation \eqref{epsilonp}
that the equation \eqref{AsymptoticIntegrandNonAdDS} implies:
\begin{equation*}
 \varepsilon,p \propto a^{2\nu-3} = a^{\sqrt{9-48\xi-4m^2/H^2}-3},
\end{equation*}
The value of the scaling exponent thus depends on the mass $m$, the Hubble parameter $H$ and 
the coupling $\xi$ but is the same for $\varepsilon$ and $p$ at variance with the classical 
case where, for $m\ne0$, $p\propto a^{-2}\varepsilon$. In the minimal coupling case $\xi=0$ 
for a very small ratio $m/H$ one gets:
\begin{equation*}
 \varepsilon, p \propto a^{-(2/3) m^2/H^2} \simeq {\rm constant}
\end{equation*}
that is, in the limit of vanishing ratio $m/H$ the energy density and the pressure become 
constant at late times. On the other hand, in the conformal case $\xi=1/6$, one has, for very 
small ratio $m/H$:
\begin{equation*}
    \varepsilon, p\propto a^{-2}
\end{equation*}
Furthermore, since the $\kk$ dependent term is the same for both $\omega_kK_k$ and $\omega_k\Gamma_k$ 
the ratio between pressure and energy density can be readily found from the equation 
\eqref{AsymptoticIntegrandNonAdDS} at late times and reads:
\begin{align*}
\dfrac{p}{\varepsilon} & \underset{t \to +\infty}{\simeq}
\dfrac{\left(1-4\xi\right)\left[-\dfrac{m^2}{H^2}+\left(\dfrac{1}{2}-\nu\right)^2+
2\left(1-6\xi\right)\right]
-2\nu\left(1-6\xi\right)}{\dfrac{m^2}{H^2}+\left(\dfrac{1}{2}-\nu\right)^2+2\left(1-\nu\right)
\left(1-6\xi\right)}
\end{align*}
with $\nu$ as in the equation \eqref{orderDeSitter}. For $\xi=0$ we have:
$$
\dfrac{p}{\varepsilon} \underset{t \to +\infty}{\simeq}
\dfrac{-\dfrac{m^2}{H^2}+\left(\dfrac{1}{2}-\nu\right)^2+2\left(1-\nu\right)}
{\dfrac{m^2}{H^2}+\left(\dfrac{1}{2}-\nu\right)^2+2\left(1-\nu\right)} =
\dfrac{-\dfrac{2m^2}{H^2}+\dfrac{9}{2}\left( 1 -\sqrt{1-\dfrac{4m^2}{9H^2}}\right)}
{\dfrac{9}{2}\left( 1 -\sqrt{1-\dfrac{4m^2}{9H^2}}\right)}
$$
For $m/H$ varying between its maximal value $3/2$ (so the order $\nu$ is real) and $0$ 
it can be readily checked that the ratio $p/\varepsilon$ varies from $0$ to its 
lowest possible value $-1$ (see equation \eqref{wlimits}) and it is thus always negative. 
Similarly, for $\xi=1/6$, we have:
$$
\dfrac{p}{\varepsilon} \underset{t \to +\infty}{\simeq}\dfrac{1}{3}
\dfrac{-\dfrac{m^2}{H^2}+\left(\dfrac{1}{2}-\nu\right)^2}
{\dfrac{m^2}{H^2}+\left(\dfrac{1}{2}-\nu\right)^2} = \dfrac{1}{3}
\dfrac{-\dfrac{2m^2}{H^2}+\dfrac{1}{2}\left( 1 -\sqrt{1-\dfrac{4m^2}{H^2}}\right)}
{\dfrac{1}{2}\left( 1 -\sqrt{1-\dfrac{4m^2}{H^2}}\right)}
$$
Again, for $m/H$ varying between its maximal value $1/2$ (so the order $\nu$ is real) and $0$ 
it can be readily checked that the ratio $p/\varepsilon$ varies from $0$ to its lowest possible
value $-1/3$ (see equation \eqref{wlimits}) and it is thus always negative.

\begin{figure}[H]
\centering
\begin{subfigure}{.45\textwidth}
  \centering
  \includegraphics[width=.99\linewidth]{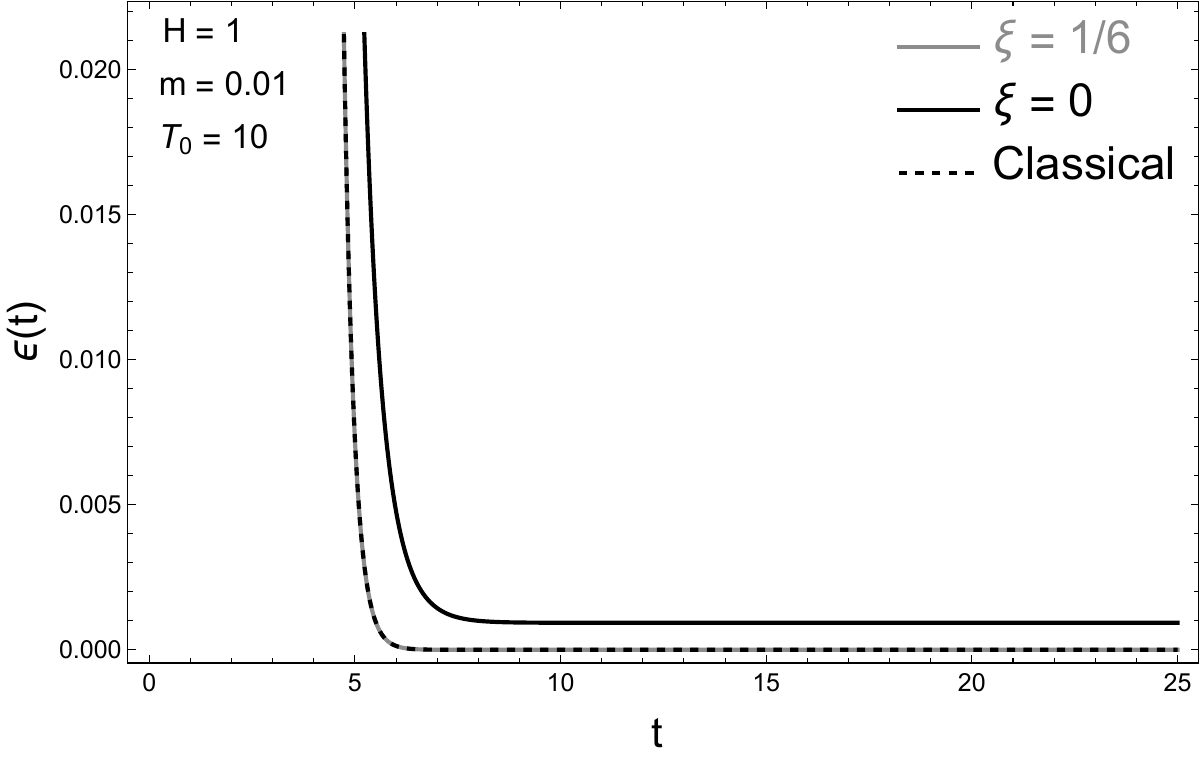}
  \caption{Energy density}
  \label{dsnad1}
\end{subfigure}%
\hspace{1cm}
\begin{subfigure}{.45\textwidth}
  \centering
  \includegraphics[width=.99\linewidth]{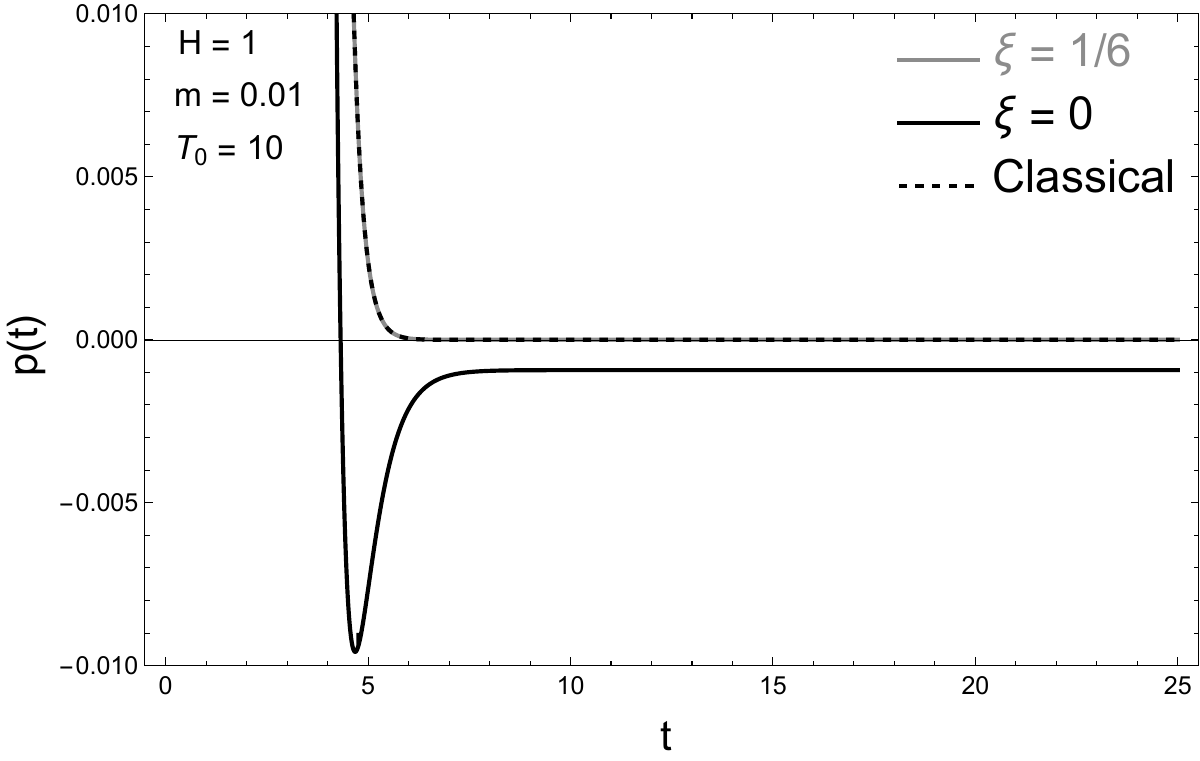}
  \caption{ Pressure}
  \label{dsnad2}
\end{subfigure}
\caption{ Energy density and pressure as a function of cosmological time for a de Sitter universe
for $H=10$, $m=0.01$ and $T_0=1$ (arbitrary units).}
\end{figure}

\begin{figure}[H]
\centering
\includegraphics[width=.55\textwidth]{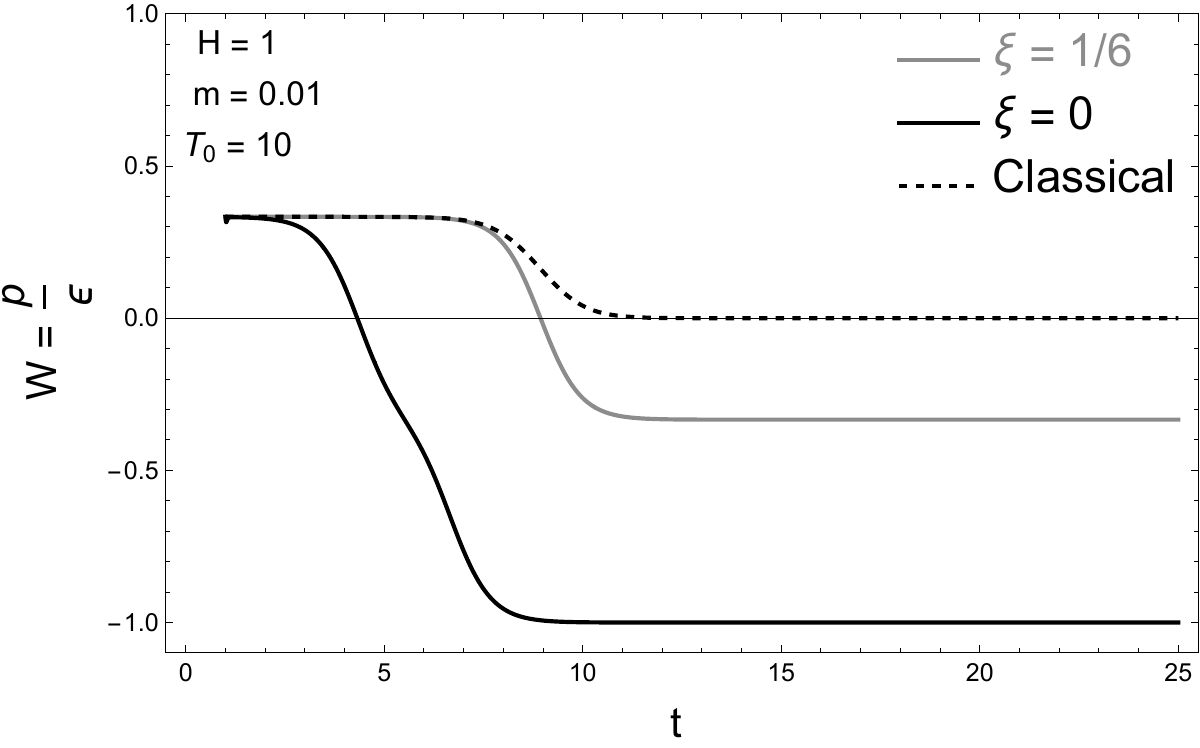}
\caption{Ratio $w$ of pressure and energy density as a function of cosmological time for a 
de Sitter universe for $H=1$, $m=0.01$ and $T_0=10$ (arbitrary units).}
\label{dsnad3}
\end{figure}

These findings imply a massive deviation from the classical relations \eqref{classical},
according to which the pressure is always positive and scales as $1/a^5$. The reason for the 
dominance of quantum corrections and why they entail a negative pressure has been discussed in depth in 
Section~\ref{sec:gen}. It should also be remarked that in the very non-adiabatic de Sitter 
expansion ($m \ll H$) for the minimally coupled scalar field energy density and pressure at
late times become constant with a ratio $w \equiv p/\varepsilon = -1$, thus mimicking a 
cosmological constant, as envisaged in Section \ref{sec:gen}.
To confirm these conclusions, we have computed the exact integrals for both $p$ and $\varepsilon$ and 
the resulting ratio $w$ for $m=0.01,T_0=1$ and $H=10$ (in arbitrary units), that is in a full 
non-adiabatic regime. The results are shown in figures~\ref{dsnad1},\ref{dsnad2} and \ref{dsnad3}. 
It can be clearly seen that, in the minimal coupling case, both the energy density and pressure 
become asymptotically constant, as expected and that the pressure has a negative asymptotic value. 
Similarly, the ratio $w = p/\varepsilon$ almost attains its minimally allowed value $-1$. 
In the conformal coupling case, energy density and pressure keep decreasing with the
expansion, but their ratio has an asymptotic limit close to $-1/3$ as expected.

\subsection{Massless case}

The massless case requires a separate discussion. For $m=0$ we have, from the eq.~\eqref{adiapards}
\begin{align*} 
 \lim_{t \to \infty} \mathcal{A}_k(t) \equiv \mathcal{A}_k(\infty) 
  =  H \,\dfrac{1}{\left[-2(1-6\xi)H^2\right]^{1/2}} = \dfrac{1}{\left[-2(1-6\xi)\right]^{1/2}}
\end{align*}
which obviously tells us that the adiabatic approximation is not a good one being $\mathcal{A}_k(\infty)$
imaginary. We thus need to solve the equation of motion \eqref{DeSitterKG} for $m=0$, which reads:
\begin{equation}\label{DeSitterKGmassless}
    \dfrac{\di^2 v_k(y)}{\di y^2}+\left[\dfrac{\kk^2}{H^2}-\dfrac{2\left(1-6\xi\right)}{y^2}\right]v_k(y)=0.
\end{equation}
The equation \eqref{DeSitterKGmassless} is again solved by a linear combination of Hankel functions 
and can be obtained by taking the $m\to0$ limit of the \eqref{SolDeSittMassNonAd} and 
\eqref{CoefficientiDeSittMassNonAd}. For $\xi \in [0,1/6]$ the order \eqref{orderDeSitter} is such that:
$$
  \dfrac{1}{2} \le \nu \le \dfrac{3}{2}
$$
For $\xi=1/6$, i.e. the conformal coupling, the equation of state is exactly $p=\varepsilon/3$ 
as it should and as it can be readily shown by setting $m=0$ and $\xi=1/6$ in the equations \eqref{funzKGAMMA} 
and \eqref{funzOmega}. On the other hand, in the minimal coupling case $\xi=0$ the equation of state 
is different. The order of the Hankel function is $\nu=3/2$ and, by using their expansion for small
argument (see Appendix A eq. \eqref{HnuRsmall}) and plugging them in \eqref{funzKGAMMA}, one obtains 
the long time limit expressions of the integrand functions of energy density and pressure:
\begin{equation}
    \begin{split}
\omega_kK_k&\sim2\left|E_k\right|^2
\left(\dfrac{H}{\kk}\right)^3\dfrac{\kk^2}{y^2},\\
\omega_k\Gamma_k&\sim2\left|E_k\right|^2
\left(\dfrac{H}{\kk}\right)^3\left(-\dfrac{\kk^2}{3y^2}\right),
    \end{split}
\end{equation}
which imply, as $y=1/a$ according to equation \eqref{ydef}, that $\varepsilon$ and $p$ scale
as $1/a^2$:
\begin{equation*}
  \varepsilon, p\propto a^{-2}.
\end{equation*}
Again, the integrands are the same function of $\kk$, hence the ratio of pressure and energy
density turns out to be a negative constant:
\begin{equation*}
    \lim_{t\to\infty}\dfrac{p}{\varepsilon}=-\dfrac{1}{3}.
\end{equation*}
Note that this result does not coincide with the limit $m\to 0$ for the minimal coupling 
$\xi=0$, where we found $p=-\varepsilon$ in the long time limit. In other words, in the 
long run of de Sitter expansion even a tiny mass is sufficient to drop the ratio $w$ from 
$-1/3$ to $-1$. This is in line with the general discussion in Section \ref{sec:gen} on the
limit of the ratio $p/\varepsilon$ for $t \to \infty$ in the minimal coupling scenario 
both for finite $m$ and $H \gg m$ and for $m=0$. The reason of the different limits is 
due to the fact that, in both the energy density and pressure expression in eq. \eqref{epsilonpt}, 
in the long run, even a tiny mass is sufficient to make the $m^2 |u_k(t)|^2$ term overwhelm 
the $\kk^2/a^2 |u_k(t)|^2$ term as $a \to \infty$ for any $\kk$.

\section{Power law expansion}
\label{sec:power}

A very general, and well behaved, class of expanding space-times is described by a scale factor 
$a(t)$ with a power law dependence on the cosmological time:
\begin{equation}\label{powerlawa}
    a(t)=\left(\dfrac{t}{t_0}\right)^\sigma,\quad\sigma>0.
\end{equation}
Note that this definition implies that $t_0$ is the time of decoupling, with $a(t_0)=1$.
The expansion is decelerated if $0<\sigma<1$ or accelerated $\sigma>1$. The case $\sigma=1$ 
corresponds to a linear expansion, which has an exact analytic solution, but we will not consider 
it in this work. Two specific cases of \eqref{powerlawa} play a special role in the standard 
cosmological models: $\sigma=1/2$, which is known as \emph{radiation\ dominated\ universe} 
and $\sigma=2/3$ known as \emph{matter\ dominated\ universe}.

The conformal time associated to the \eqref{powerlawa} is remarkably different for decelerated 
and accelerated expansions. In the former case, and the linear expansion as well, the conformal
time is infinite for $t \to +\infty$, whereas in the latter case, the conformal time $\eta$ is 
finite in the same limit and so there is a cosmological horizon $\eta_\infty=t_0/(\sigma-1)$. 
The scale factor as a function of conformal time reads:
\begin{equation}\label{conformalscalespower}
    \begin{split}
a(\eta)=\left(1+h\eta\right)^\frac{\sigma}{1-\sigma},\quad  h\equiv\dfrac{1-\sigma}{ t_0},
 \ 0<\sigma<1 &\qquad \mbox{decelerated\ expansion},\\
a(\eta)=\dfrac{1}{\left(1-h\eta\right)^{\frac{\sigma}{\sigma-1}}},\quad h \equiv \dfrac{\sigma-1}{ t_0} 
= \eta^{-1}_\infty,\ 
\sigma>1 &\qquad \mbox{accelerated\ expansion},\\
a(\eta)=\e^{h\eta},\quad h=t^{-1}_0,\ \sigma=1 &\qquad\mbox{linear\ expansion}.
    \end{split}
\end{equation}
It is useful to define the parameter $\delta_\sigma$:
\begin{equation}\label{deltaSigma}
   \delta_\sigma\equiv\left(1-6\xi\right)\dfrac{\sigma\left(2\sigma-1\right)}{\left(1-\sigma\right)^2},
\end{equation}
and the adimensional variable $y$:
\begin{equation}\label{yeta}
    y=1\pm h\eta,
\end{equation}
where the plus applies to the decelerated expansion and the minus to the accelerated expansion.

We will cope with the massive and massless case separately.

\subsection{Massive case}

In the massive case, the adiabatic parameter \eqref{adiapar} can be written as:
\begin{equation*}
    \mathcal{A}_k=\dfrac{h}{m}\dfrac{\sigma}{1-\sigma}\dfrac{1}{y^\frac{\sigma}{1-\sigma}}
    \dfrac{1+\dfrac{h^2}{m^2}\dfrac{1-\sigma}{\sigma}\dfrac{\delta_\sigma}{y^{\frac{2}{1-\sigma}}}}
    {\left[1+\dfrac{\kk^2}{m^2}\dfrac{1}{y^{\frac{2\sigma}{1-\sigma}}}-
    \dfrac{h^2}{m^2}\dfrac{\delta_\sigma}{y^\frac{2}{1-\sigma}}\right]^{3/2}}.
\end{equation*}
The late time limit corresponds to $y\to\infty$ in the decelerated case and $y \to 0$ in the accelerated
one. In both cases, we have, for a massive field: 
$$
   \lim_{t \to +\infty} \mathcal{A}_k(t) = 0 
$$
that is, for a general power law expansion the field will evolve adiabatically regardless 
of acceleration at late times. On the other hand, at the decoupling $y=1$ and $\eta=0$, 
so, for $\sigma \ne 1$:
\begin{equation*}
    \mathcal{A}_k(0) = \dfrac{h}{m}\dfrac{\sigma}{1-\sigma} 
    \dfrac{1+\dfrac{h^2}{m^2}\dfrac{1-\sigma}{\sigma}\delta_\sigma}
    {\left[1+\dfrac{\kk^2}{m^2}-\dfrac{h^2}{m^2}\delta_\sigma\right]^{3/2}}.
\end{equation*}
From this equation it appears that the adiabaticity at the decoupling is mostly driven 
by the ratio $h/m$, that is:
$$
  \dfrac{h}{m}= \dfrac{|\sigma -1|}{m \, t_0}  \; .
$$
If $h/m \ll1$, that is $m t_0 \gg 1$, the field is in the adiabatic regime throughout; if, on
the other hand, $h/m \gtrsim 1$, that is $ m \, t_0 \lesssim 1$ the field is not in the adiabatic 
regime at the decoupling and we can expect interesting effects on pressure and energy density
analogous to those found in the de Sitter universe.

For a general power law expanding universe we could not find an exact analytic solution of the equation 
\eqref{kg2}. Thus, we worked out a numerical solution of the differential equation by using Mathematica
\cite{Mathematica}. In this Section, we report the obtained numerical results for the following cases:
\begin{itemize}
    \item A radiaton dominated universe: $\sigma=1/2$,
    \item A matter dominated universe: $\sigma=2/3$,
    \item A quadratic accelerated universe: $\sigma=2$
\end{itemize}

\subsubsection{Decelerated expansion: $\sigma <1$}

The differential equation \eqref{kg2} reads:
\begin{equation}  v''_k+\left[\kk^2+m^2\left(1+h\eta\right)^{\frac{2\sigma}{1-\sigma}}-
    \dfrac{h^2\delta_\sigma}{\left(1+h\eta\right)^2}\right]v_k=0,
\end{equation}
with $\delta_\sigma$ given in the eq. \eqref{deltaSigma}. The energy density and pressure have been
obtained by solving the differential equation numerically \cite{Mathematica} for $\sigma=2/3$
for the adiabatic case with $h/m = 0.2$ and for $\sigma=1/2$ for a non adiabatic case with $h/m= 10$.
\begin{figure}[H]
\centering
\begin{subfigure}{.45\textwidth}
  \centering
  \includegraphics[width=.99\linewidth]{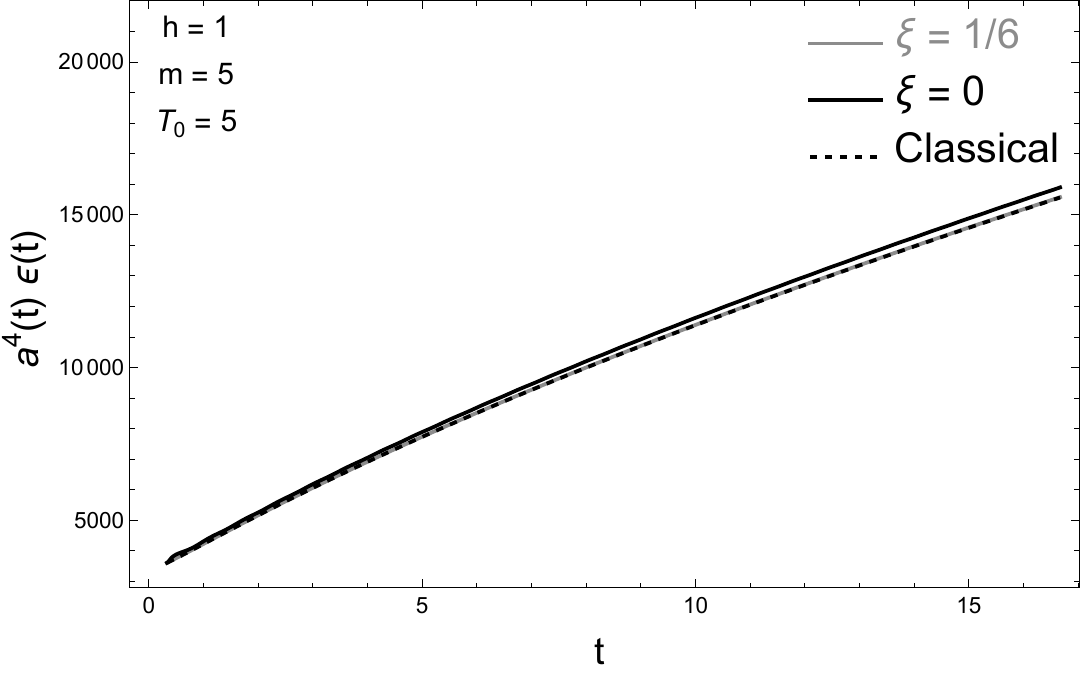}
  \caption{Energy density}
  \label{mdad1}
\end{subfigure}%
\hspace{1cm}
\begin{subfigure}{.45\textwidth}
  \centering
  \includegraphics[width=.99\linewidth]{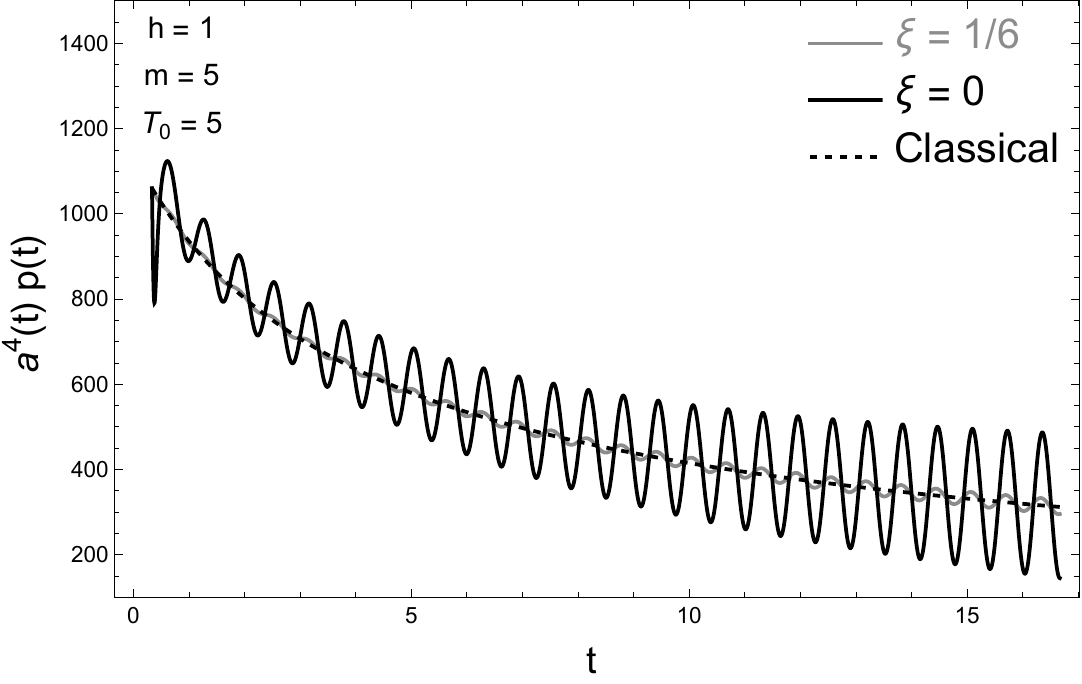}
  \caption{Pressure}
  \label{mdad2}
\end{subfigure}
\caption{Energy density and pressure as a function of cosmological time for $a(t)=(t/t_0)^{2/3}$
for $h=1$, $m=5$ and $T_0=5$ (arbitrary units).}
\end{figure}
\begin{figure}[H]
\centering
\includegraphics[width=.55\textwidth]{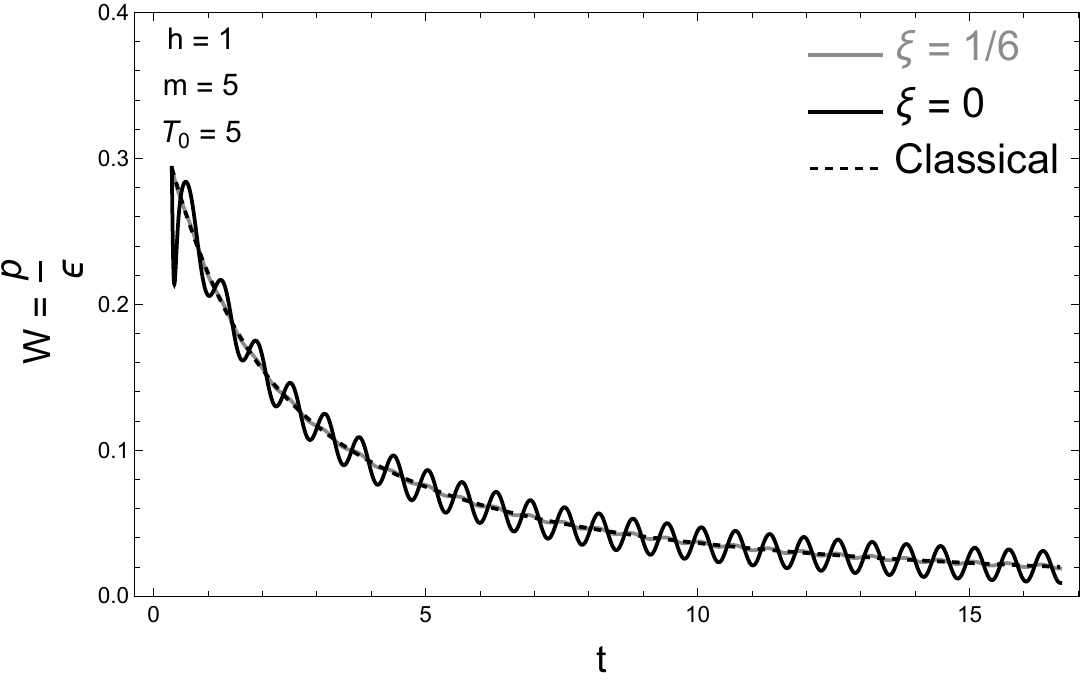}
\caption{Ratio of pressure and energy density as a function of cosmological time for 
$a(t)=(t/t_0)^{2/3}$ for $h=1$, $m=5$ and $T_0=5$ (arbitrary units).}
\label{mdad3}
\end{figure}
\begin{figure}[h!]
\centering
\begin{subfigure}{.45\textwidth}
  \centering
  \includegraphics[width=.99\linewidth]{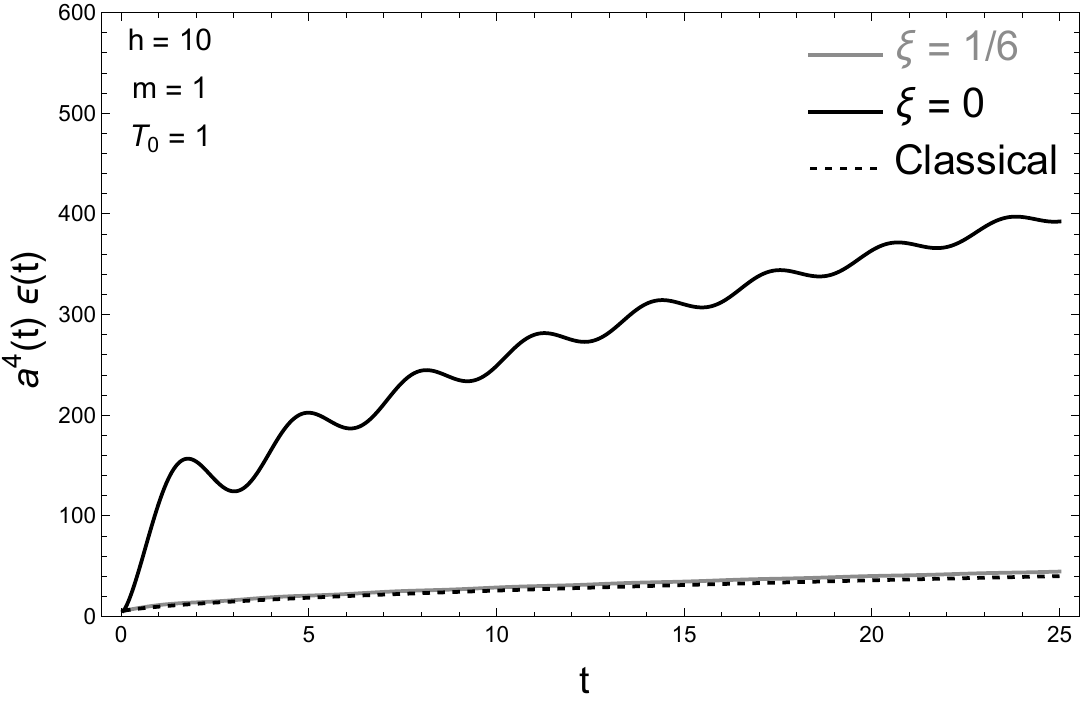}
  \caption{Energy density}
  \label{rdnad1}
\end{subfigure}%
\hspace{1cm}
\begin{subfigure}{.45\textwidth}
  \centering
  \includegraphics[width=.99\linewidth]{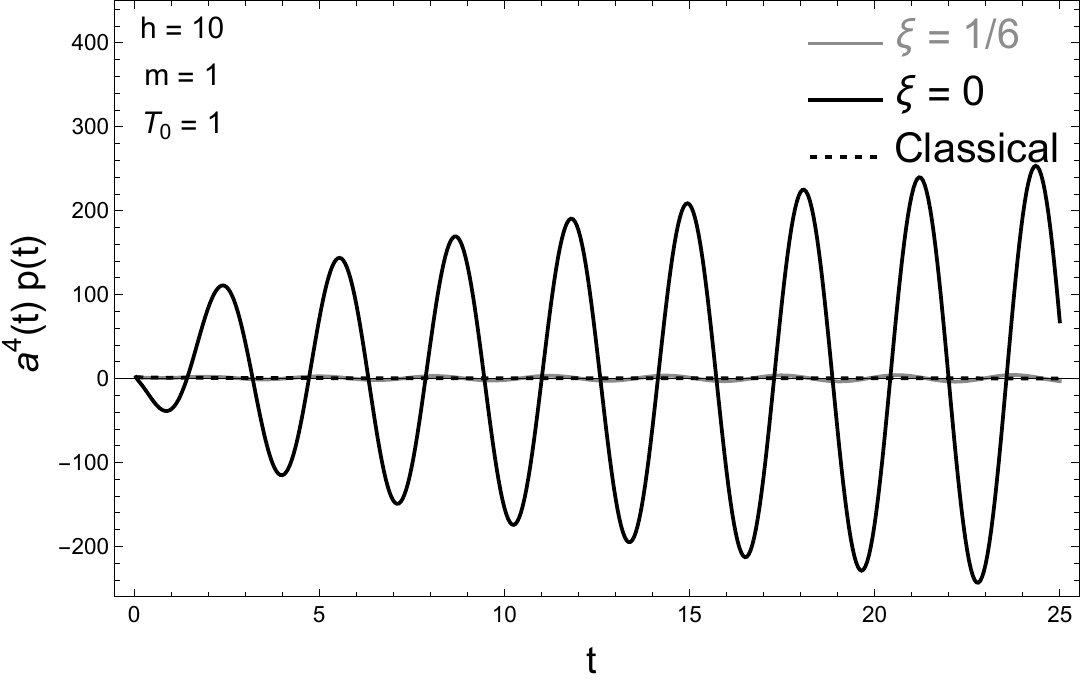}
  \caption{Pressure}
  \label{rdnad2}
\end{subfigure}
\caption{Energy density and pressure as a function of cosmological time for $a(t)=(t/t_0)^{1/2}$
for $h=10$, $m=1$ and $T_0=1$ (arbitrary units).}
\end{figure}
\begin{figure}[h!]
\centering
\includegraphics[width=.55\textwidth]{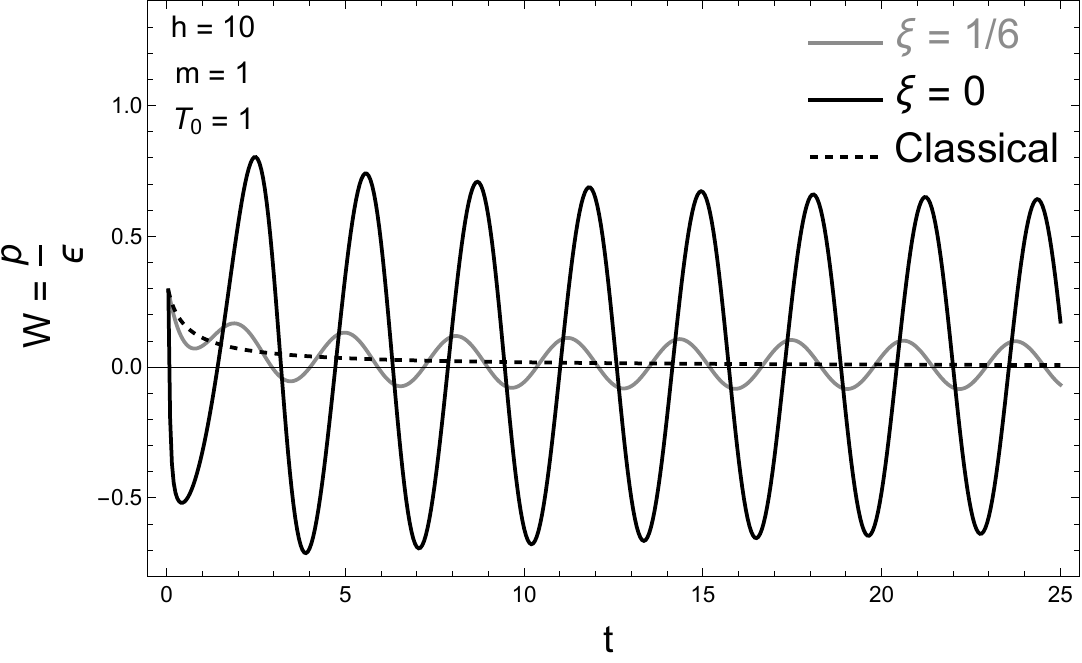}
\caption{Ratio of pressure and energy density as a function of cosmological time for 
$a(t)=(t/t_0)^{1/2}$ for $h=10$, $m=1$ and $T_0=1$ (arbitrary units).}
\label{rdnad3}
\end{figure}
While in the full adiabatic case, with $h/m =0.2$ one can see that (see figures~\ref{mdad1}, 
\ref{mdad2} and \eqref{mdad3}), as expected, quantum corrections boil down to small oscillations 
around the classical values, in the non-adiabatic decoupling case, with $h = 1/2t_0 \gg m$, 
for $\xi=0$ (see figures~\ref{rdnad1}, \ref{rdnad2} and \eqref{rdnad3}) pressure becomes negative 
and the ratio $w$ attains significantly negative values after decoupling before starting 
oscillating with a frequency proportional to the mass and a large amplitude.

\subsubsection{Accelerated expansion: $\sigma > 1$}

\begin{figure}[H]
\centering
\begin{subfigure}{.45\textwidth}
  \centering
  \includegraphics[width=.99\linewidth]{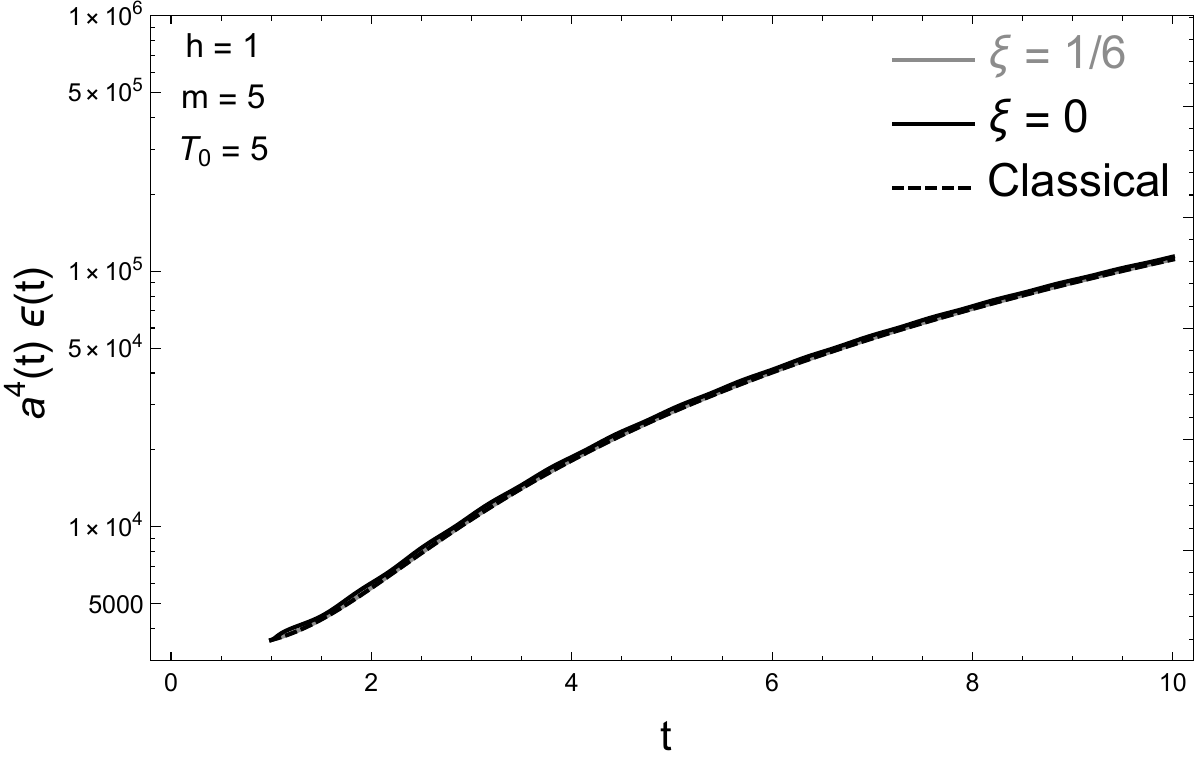}
  \caption{Energy density}
  \label{accad1}
\end{subfigure}%
\hspace{1cm}
\begin{subfigure}{.45\textwidth}
  \centering
  \includegraphics[width=.99\linewidth]{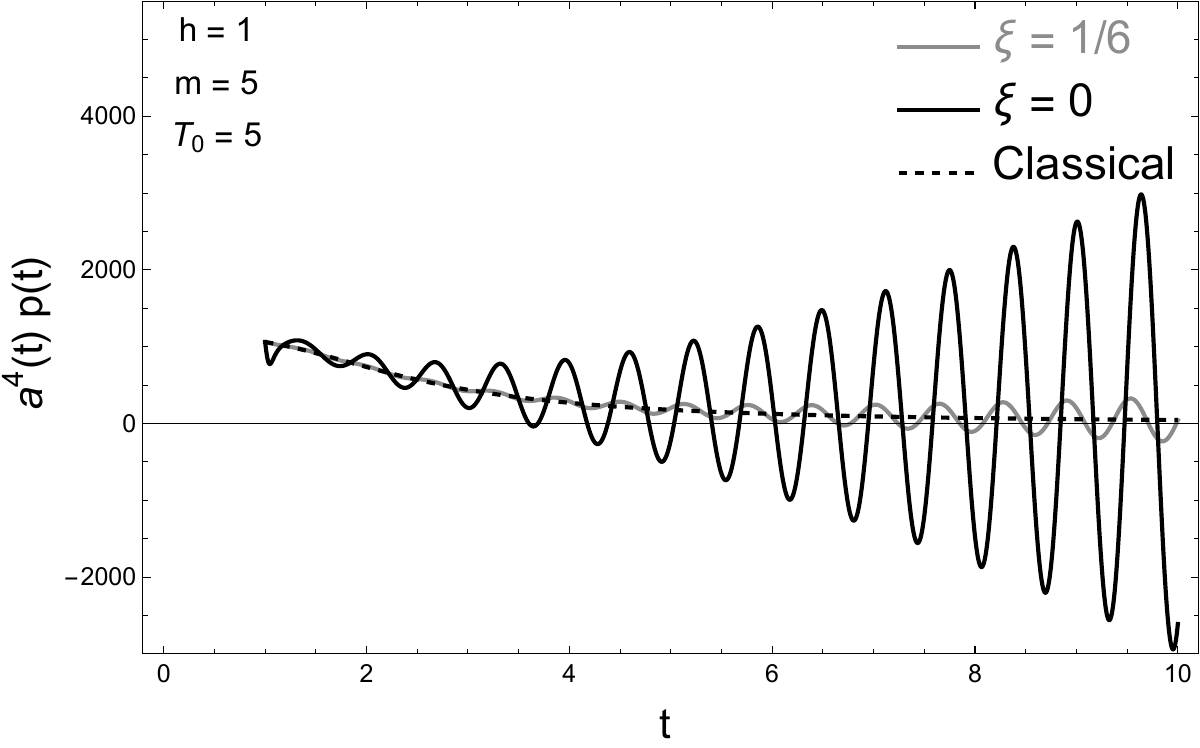}
  \caption{Pressure}
  \label{accad2}
\end{subfigure}
\caption{Energy density and pressure as a function of cosmological time for $a(t)=(t/t_0)^2$
for $h=1$, $m=5$ and $T_0=5$ (arbitrary units).}
\end{figure}

\begin{figure}[H]
\centering
\includegraphics[width=.55\textwidth]{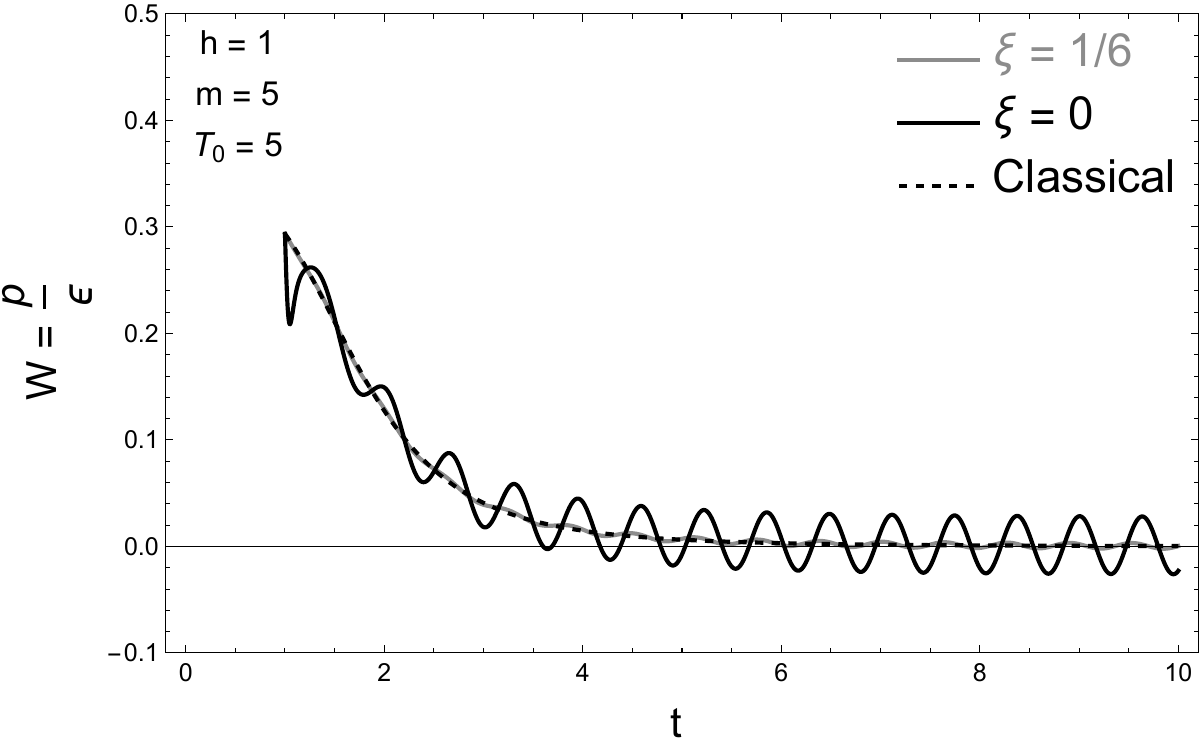}
\caption{Ratio of pressure and energy density as a function of cosmological time for 
$a(t)=(t/t_0)^2$ for $h=1$, $m=5$ and $T_0=5$ (arbitrary units).}
\label{accad3}
\end{figure}

\begin{figure}[H]
\centering
\begin{subfigure}{.45\textwidth}
  \centering
  \includegraphics[width=.99\linewidth]{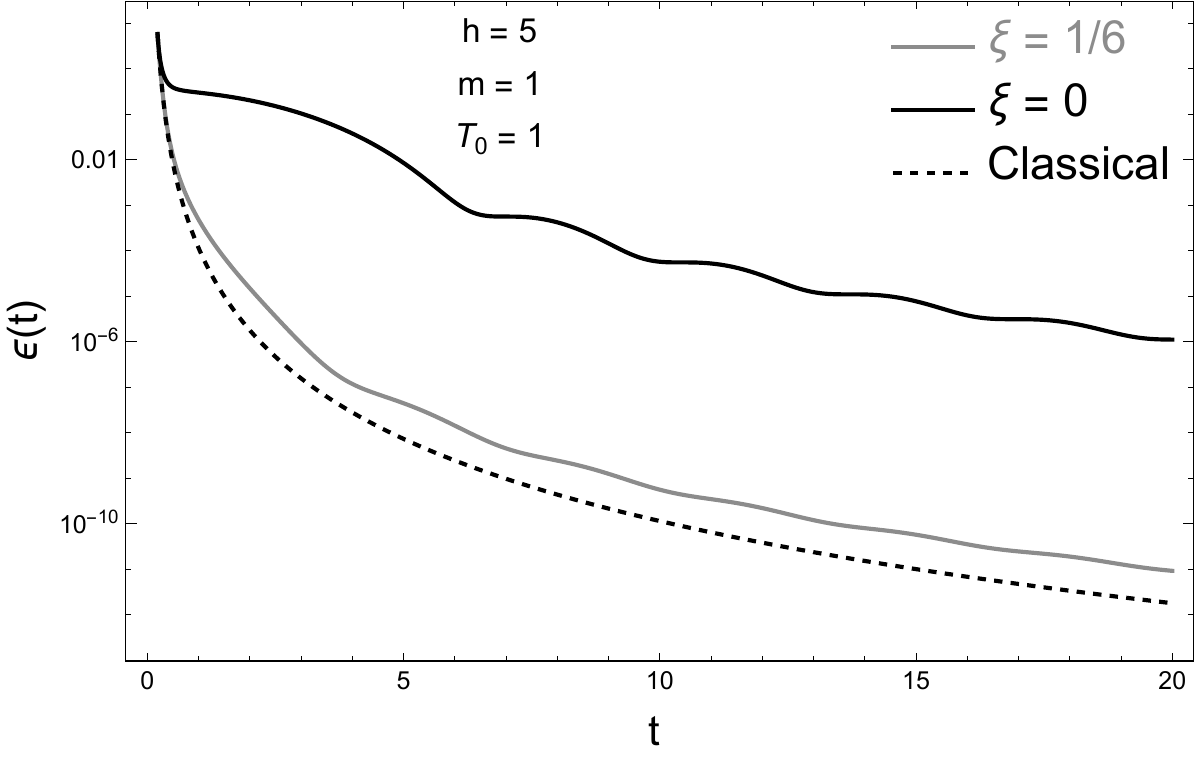}
  \caption{Energy density}
  \label{accnad1}
\end{subfigure}%
\hspace{1cm}
\begin{subfigure}{.45\textwidth}
  \centering
  \includegraphics[width=.99\linewidth]{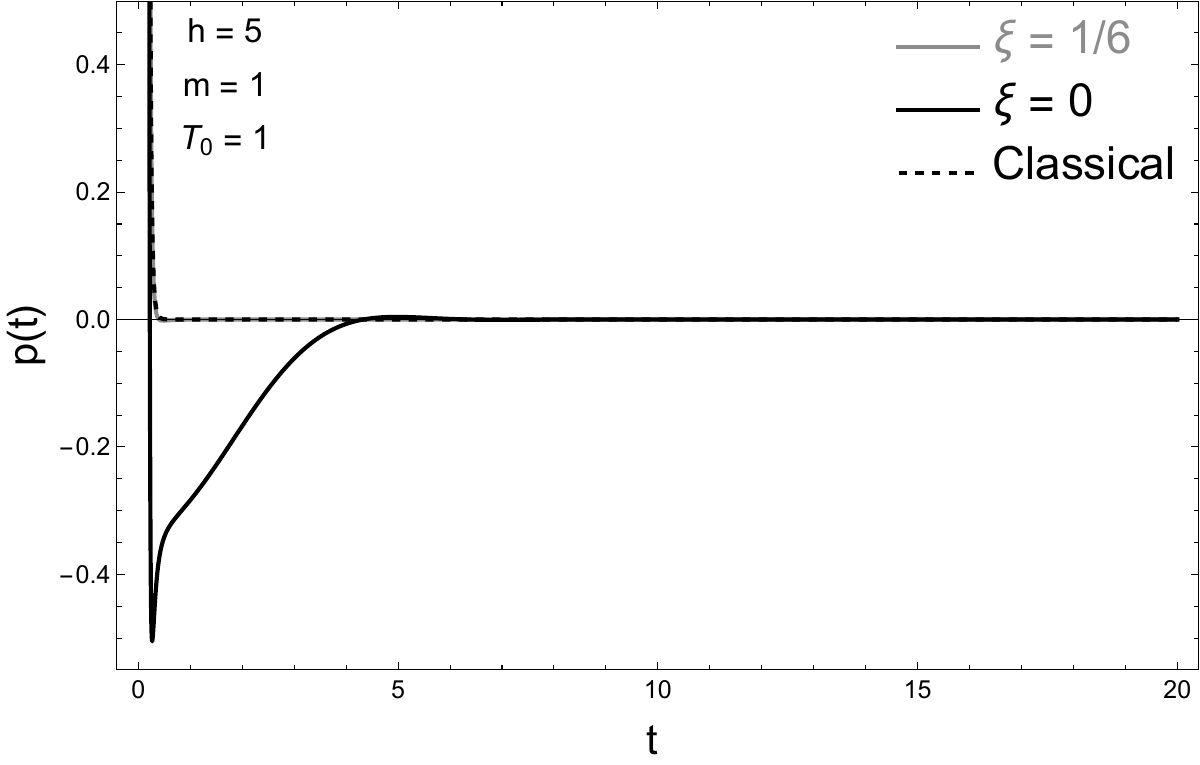}
  \caption{Pressure}
  \label{accnad2}
\end{subfigure}
\caption{Energy density and pressure as a function of cosmological time for $a(t)=(t/t_0)^2$
for $h=5$, $m=1$ and $T_0=1$ (arbitrary units).}
\end{figure}

\begin{figure}[H]
\centering
\includegraphics[width=.55\textwidth]{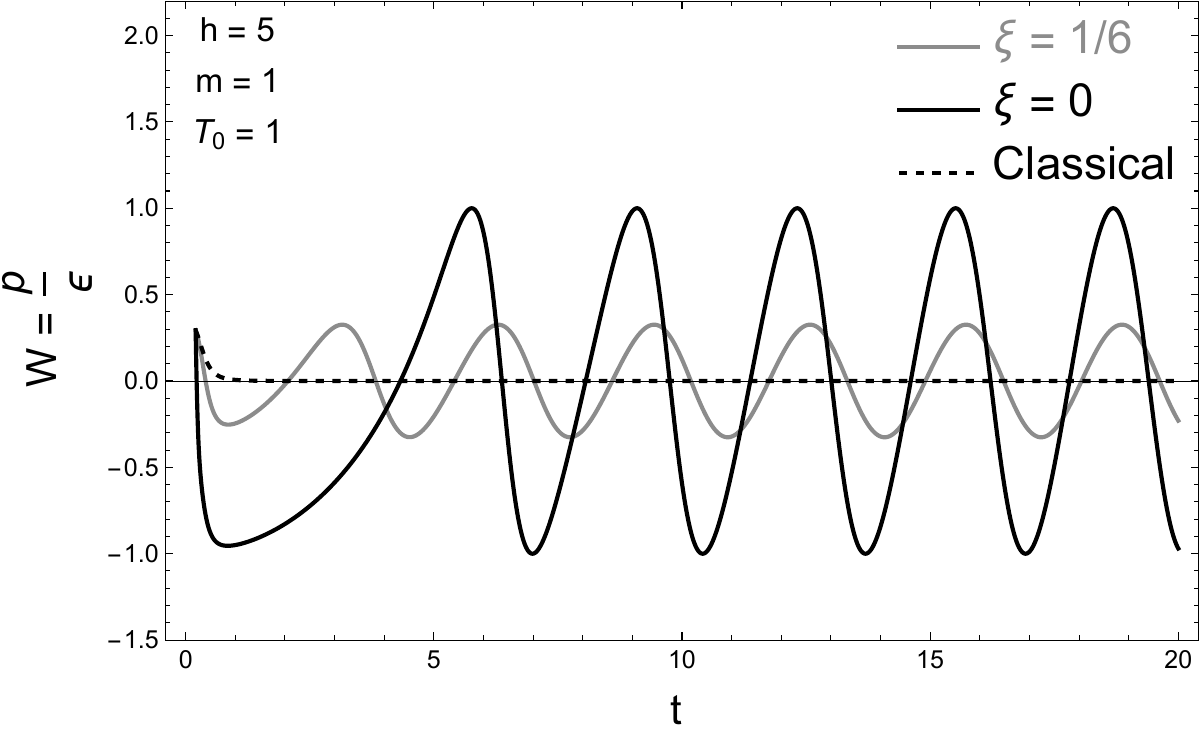}
\caption{Ratio of pressure and energy density as a function of cosmological time for 
$a(t)=(t/t_0)^2$ for $h=5$, $m=1$ and $T_0=1$ (arbitrary units).}
\label{accnad3}
\end{figure}

In this case the differential equation \eqref{kg2} reads:
\begin{equation}
v''_k+\left[\kk^2+\dfrac{m^2}{\left(1-h\eta\right)^{\frac{2\sigma}{\sigma-1}}}-
\dfrac{h^2\delta_\sigma}{\left(1-h\eta\right)^2}\right]v_k=0,
\end{equation}
The energy density and pressure have been obtained by solving the differential equation numerically 
\cite{Mathematica} for $\sigma=2$ and the results are shown in figures~\ref{accad1}, \ref{accad2} 
and \eqref{accad3} for the adiabatic case ($h/m = 0.2$) and in figures \ref{accnad1}, \ref{accnad2} 
and \ref{accnad3} for a non adiabatic case ($h/m= 5$). Again, like in the non-accelerated case,
for an adiabatic decoupling with $h/m=0.2$ quantum corrections come down to oscillations 
around the classical values, whereas in the non-adiabatic decoupling case, with $h/m=5$, 
the ratio $w$ attains a large negative value after decoupling before getting back to
positive values, thereafter oscillating with a frequency proportional to the mass and a large 
amplitude.

\subsection{Massless case}

The massless case is a peculiar one in that for $\xi=1/6$ conformal invariance implies that the
equation of state is always $p=(1/3)\varepsilon$. A more interesting case then occurs when $\xi \ne 1/6$ 
and particularly in the minimal coupling case $\xi=0$; we will henceforth focus on the
latter.

For a massless field the differential equations \eqref{kg2} can be solved analytically and the
solutions are again Hankel functions of real order. However, there is a noteworthy difference between 
decelerated and accelerated expansion. In the former, the asymptotic evolution turns out 
to be adiabatic regardless of the decoupling conditions whereas in the latter the evolution at 
late times is non-adiabatic. This can be understood more easily by referring to the discussion 
in Section \ref{sec:gen}. Since:
$$
  \dot a = \sigma \frac{t^{\sigma-1}}{t_0}
$$
we can conclude that if $\sigma < 1$, in the long run $\dot a(t) \to 0$; thus $\dot a (t) \ll T_0$
and an oscillatory behaviour of the mode functions is expected, i.e an adiabatic regime. 
More specifically, all the derivatives of the scale factor can be neglected and so one can readily
show by using the formulae \eqref{epsilonp} and \eqref{funzKGAMMA} that the ratio $w = p/\varepsilon$
asymptotically becomes $1/3$, the classical value. Conversely, if $\sigma > 1$, $\dot a \to \infty$ 
in the long run and since $\dot a \gg T_0$ one expects a slowly decreasing exponential solution 
\eqref{expsol} for the mode, whence an equation of state $p=-(1/3) \varepsilon$.

This analysis is confirmed by the calculation of the adiabatic parameter \eqref{adiapar} 
for $m=0$:
\begin{equation*}
    \mathcal{A}_k=\dfrac{1-6\xi}{2\left[\kk^2-\left(1-6\xi\right)\dfrac{a''}{a}\right]^{3/2}}
    \left(\dfrac{a''a'}{a^2}-\dfrac{a'''}{a}\right),
\end{equation*}
which, by using the \eqref{conformalscalespower} and the \eqref{deltaSigma}, can be written as:
\begin{equation}\label{AdiabaticParameterMasslessPoweLaw}
\mathcal{A}_k=\dfrac{\delta_\sigma}{\left[\dfrac{\kk^2y^2}{h^2}-\delta_\sigma\right]^{3/2}}
\end{equation}
The adiabatic parameter in the long run strongly depends on the trend of $y$ as a function of 
cosmological time. On the other hand, for $\eta=0$ we have $y=1$ anyway and the 
adiabatic parameter \eqref{AdiabaticParameterMasslessPoweLaw} at the decoupling becomes:
\be\label{AdiabaticParameterMasslessPoweLawdec}
\mathcal{A}_k=\dfrac{\delta_\sigma}{\left[\dfrac{\kk^2}{h^2}-\delta_\sigma\right]^{3/2}}
\ee
We are now going to present the results for the decelerated and accelerated expansions 
separately.

\subsubsection{Decelerated expansion}

For a decelerated expansion $0<\sigma<1$ and $y=1+h\eta$, according to the equation \eqref{yeta}.
At large times $t\to +\infty$ one has $y\to +\infty$, and the adiabatic parameter in 
\eqref{AdiabaticParameterMasslessPoweLaw} is apparently very small. Yet, there is a subtlety concerned 
with the sign of $\delta_\sigma$. If $0 < \sigma < 1/2$  the coefficient $\delta_\sigma$ in the 
eq.~\eqref{deltaSigma} is negative, hence the radicand in the denominator of \eqref{AdiabaticParameterMasslessPoweLaw} 
is positive. Therefore, for $y\to +\infty$ the dependence of the adiabatic parameter on $y$ is
as follows:
\begin{equation*}
    \mathcal{A}_k \underset{y \to +\infty}{\simeq} \delta_\sigma \frac{h^3}{\kk^3} \dfrac{1}{y^3}   
\end{equation*}
On the other hand, for $1/2 < \sigma < 1$, $\delta_\sigma$ is positive and it may well happen that 
the radicand in the denominator of \eqref{AdiabaticParameterMasslessPoweLaw} stays negative even at 
very late times if the ratio $\kk/h$ is small, that is if:
\begin{equation*}
    \dfrac{\kk \, y}{h} = \dfrac{\kk \, t_0 \, y}{1-\sigma} \lesssim \delta_\sigma^{1/2},
\end{equation*}
For such modes, the adiabatic approximation at long times is not possible. Since the momentum 
scale is dictated by $T_0$ the smaller the ratio $T_0/h$, the longer the deviation from 
the adiabatic approximation. Similar arguments apply at the decoupling time, 
when the adiabatic parameter is given by eq.~\eqref{AdiabaticParameterMasslessPoweLawdec}.

The differential equation \eqref{kg2} reads:
\begin{equation*}
    \dfrac{\di^2v_k}{\di y^2}+\left[\dfrac{\kk^2}{h^2}-\dfrac{\delta_\sigma}{y^2}\right]v_k=0.
\end{equation*}
with $\delta_\sigma$ given by \eqref{deltaSigma}. The general solution is a combination of Hankel 
functions:
\begin{equation}
    v_k(y)=C_k\sqrt{\dfrac{\pi}{4h}}\sqrt{y} \, \hdue\left(\dfrac{\kk y}{h}\right)+
    D_k\sqrt{\dfrac{\pi}{4h}}\sqrt{y} \, \huno\left(\dfrac{\kk y}{h}\right),
\end{equation}
where the order $\nu$ is given by
\begin{equation*}
    \nu=\dfrac{1}{2}\sqrt{1+4\delta_\sigma}.
\end{equation*}
The order $\nu$ is always positive for $0 < \sigma < 1$ and, particularly, is $< 1/2$ for 
$0<\sigma\le 1/2$ whereas it is $>1/2$ if $\sigma >1/2$. The two coefficients $C_k$ and $D_k$ are, 
as usual, determined by the initial conditions \eqref{condin} and turn out to be:
\begin{equation}
\begin{cases}
C_k=\sqrt{\dfrac{2h}{\pi\omega_\xi(0)}}\dfrac{1}{\hdue\left(k/h\right)}+\sqrt{\dfrac{\pi\omega_\xi(0)}{8h}}\huno\left(\kk/h\right)\left[1+\dfrac{\ii h}{2\omega_\xi(0)}\left(-1+\dfrac{2\left(1-6\xi\right)\sigma}
{1-\sigma}-2\nu+\dfrac{2\kk}{h}\dfrac{\hduepiu\left(\kk/h\right)}{\hdue\left(\kk/h\right)}
\right)\right],\\
D_k=-\sqrt{\dfrac{\pi\omega_\xi(0)}{8h}}\hdue\left(\kk/h\right)\left[1+\dfrac{\ii h}{2\omega_\xi(0)}\left(-1+\dfrac{2\left(1-6\xi\right)\sigma}
{1-\sigma}-2\nu+\dfrac{2\kk}{h}\dfrac{\hduepiu\left(\kk/h\right)}{\hdue\left(\kk/h\right)}
\right)\right],
\end{cases}
\end{equation}
It can be checked that they fulfill the normalization condition $\left|C_k\right|^2-\left|D_k\right|^2=1$.

\begin{figure}[H]
\centering
\begin{subfigure}{.45\textwidth}
  \centering
  \includegraphics[width=.99\linewidth]{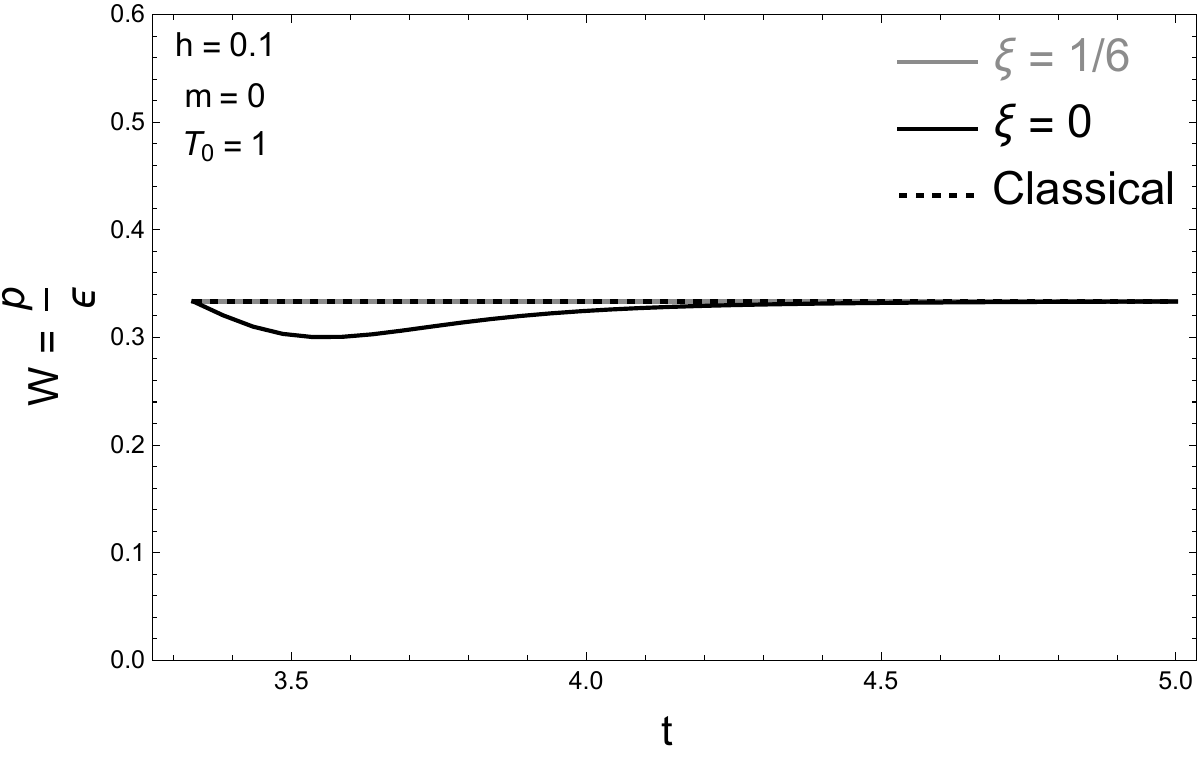}
  \caption{Ratio of pressure and energy density as a function of time for 
$a(t)=(t/t_0)^{2/3}$ for $h=0.1$ and $T_0=1$ (arbitrary units) for a massless field.}
  \label{zeromass1}
\end{subfigure}
\hspace{1cm}
\begin{subfigure}{.45\textwidth}
  \centering
  \includegraphics[width=.99\linewidth]{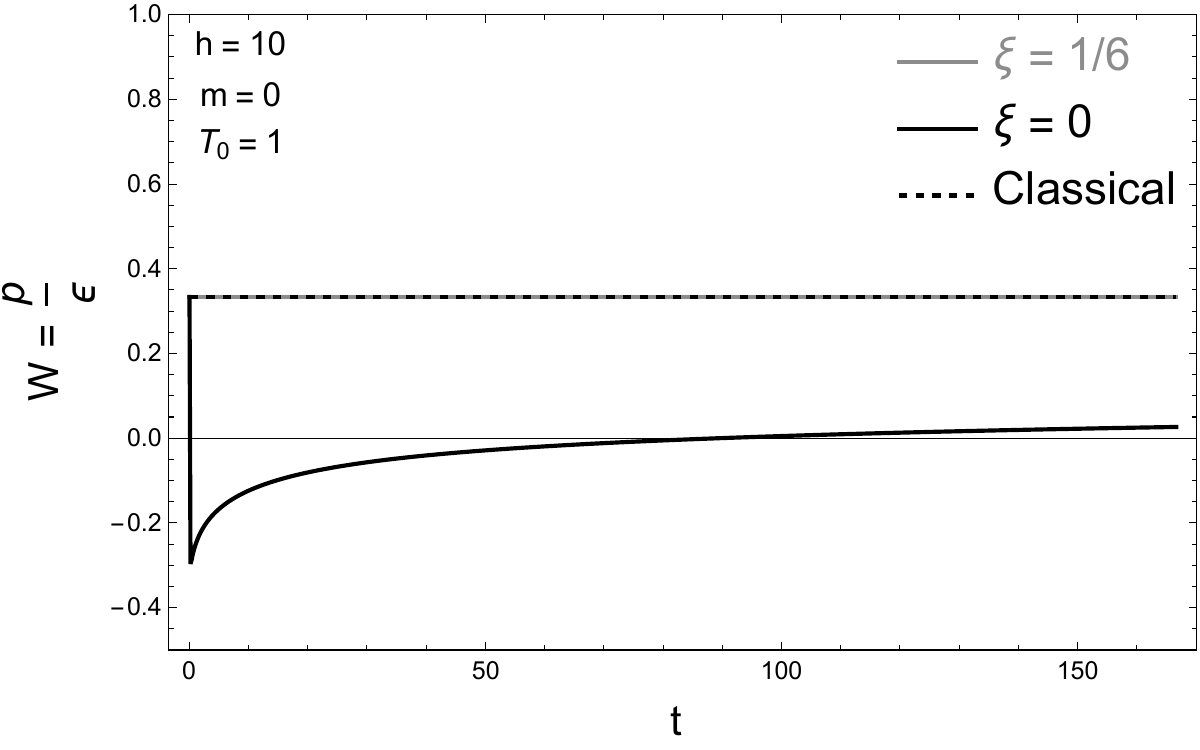}
  \caption{Ratio of pressure and energy density as a function of time for 
$a(t)=(t/t_0)^{2/3}$ for $h=10$ and $T_0=1$ (arbitrary units) for a massless field.}
  \label{zeromass2}
\end{subfigure}
\caption{}
\end{figure}

We have calculated the ratio pressure over energy density for $\sigma=2/3$ for two cases: 
$h=0.1, T_0=1$ and $h=10, T_0=1$ (arbitrary units), see figures \ref{zeromass1} and \ref{zeromass2}.
In the first case, the adiabatic approximation works well at finite time for most modes, as has 
been discussed, and it can be seen that after a relatively short time the ratio converges to 
its classical limit. Conversely, for a non-adiabatic decoupling with $h/T_0=10$ the ratio 
attains almost the minimal allowed value of $-1/3$, then it very slowly rises towards the 
classical value of $1/3$.

\subsubsection{Accelerated expansion}

For an accelerated expansion with $\sigma>1$, at large times $t \to +\infty$ 
one has $y\to0$ while from equation \eqref{deltaSigma} $\delta_\sigma>0$; thus, the adiabatic 
parameter in eq.~\eqref{AdiabaticParameterMasslessPoweLaw} becomes imaginary and so 
the adiabatic approximation is not asymtpotically valid, as expected. Nevertheless, if 
$T_0 \gg h$, at the decoupling the adiabatic parameter in \eqref{AdiabaticParameterMasslessPoweLawdec} 
is small and the adiabatic approximation is likely to work well for some time. Otherwise, 
unlike the massive case, in an accelerated expanding universe the evolution of the massless
field is always non-adiabatic. For an accelerated power law expansion the differential 
equation \eqref{kg2} reads:
\begin{equation}\label{masslessEQACC}
    v''_k(\eta)+\left[\kk^2-\dfrac{\delta_\sigma h^2}{\left(1-h\eta\right)^2}\right]v_k(\eta)=0.
\end{equation}
As it has been discussed, the $y$ parameter can now be defined as:
\begin{equation*}
    y=1-h\eta,\quad \di y=-h\di\eta,
\end{equation*}
and in the the late time limit $\eta\to h^{-1}$, corresponding to $y\to0$. The normalized solution 
is proportional to the Hankel function $\huno$:
\begin{equation}\label{masslessEQACCsol}
    v_k(y)=C_k\sqrt{\dfrac{\pi}{4H}}\sqrt{y}\,\huno\left(\dfrac{\kk y}{h}\right)+
    B_k\sqrt{\dfrac{\pi}{4H}}\sqrt{y}\, \hdue\left(\dfrac{\kk y}{h}\right),
\end{equation}
with order $\nu$ given by
\begin{equation*}
    \nu=\dfrac{1}{2}\sqrt{4\delta_\sigma+1},
\end{equation*}
which is always real and positive for $\sigma>1$ and $\xi\in\left[0,1/6\right)$. The normalized
coefficients now read:
\begin{equation}\label{CoeffDecMassless}
\begin{cases}
C_k=\sqrt{\dfrac{2h}{\pi\omega_\xi(0)}}\dfrac{1}{\huno\left(\kk/h\right)}+
\sqrt{\dfrac{\pi\omega_\xi(0)}{8h}}\hdue\left(\kk/h\right)
\left[1+\dfrac{\ii h}{2\omega_\xi(0)}\left(1+\dfrac{2\left(1-6\xi\right)\sigma}{\sigma-1}+2\nu-2\dfrac{\kk}{h}
\dfrac{\hunopiu\left(\kk/h\right)}{\huno\left(\kk/h\right)}\right)\right],\\
D_k=-\sqrt{\dfrac{\pi\omega_\xi(0)}{8h}}\huno\left(\kk/h\right)
\left[1+\dfrac{\ii h}{2\omega_\xi(0)}\left(1+\dfrac{2\left(1-6\xi\right)\sigma}{\sigma-1}+2\nu-2\dfrac{\kk}{h}
\dfrac{\hunopiu\left(\kk/h\right)}{\huno\left(\kk/h\right)}\right)\right].
\end{cases}
\end{equation}
It is worth pointing out that in the limit $\sigma\to\infty$ the equation \eqref{masslessEQACC}, 
its solution \eqref{masslessEQACCsol} and the coefficients \eqref{CoeffDecMassless} come down 
to those obtained for the massless field in the de Sitter universe, with $h\to H$.

\begin{figure}[H]
\centering
\begin{subfigure}{.45\textwidth}
  \centering
  \includegraphics[width=.99\linewidth]{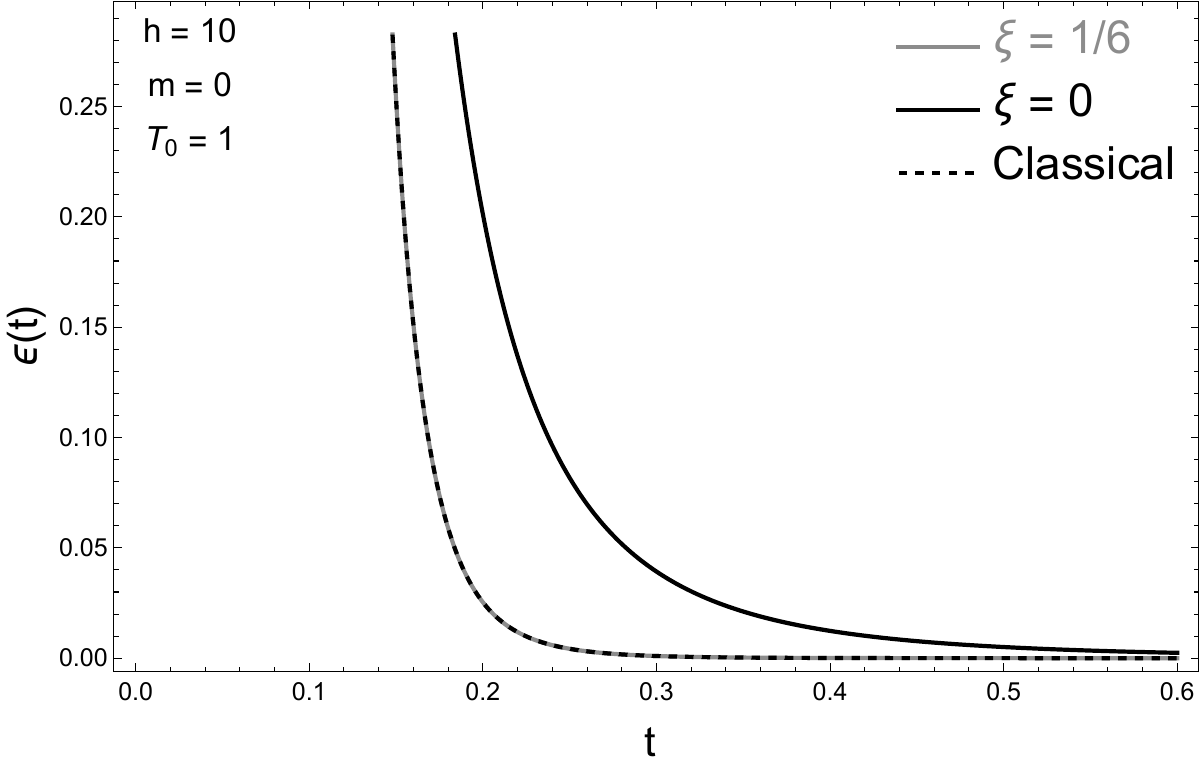}
  \caption{Energy density as a function of time for 
$a(t)=(t/t_0)^2$ for $h=10$ and $T_0=1$ (arbitrary units) for a massless field.}
  \label{zeromassend}
\end{subfigure}%
\hspace{1cm}
\begin{subfigure}{.45\textwidth}
  \centering
  \includegraphics[width=.99\linewidth]{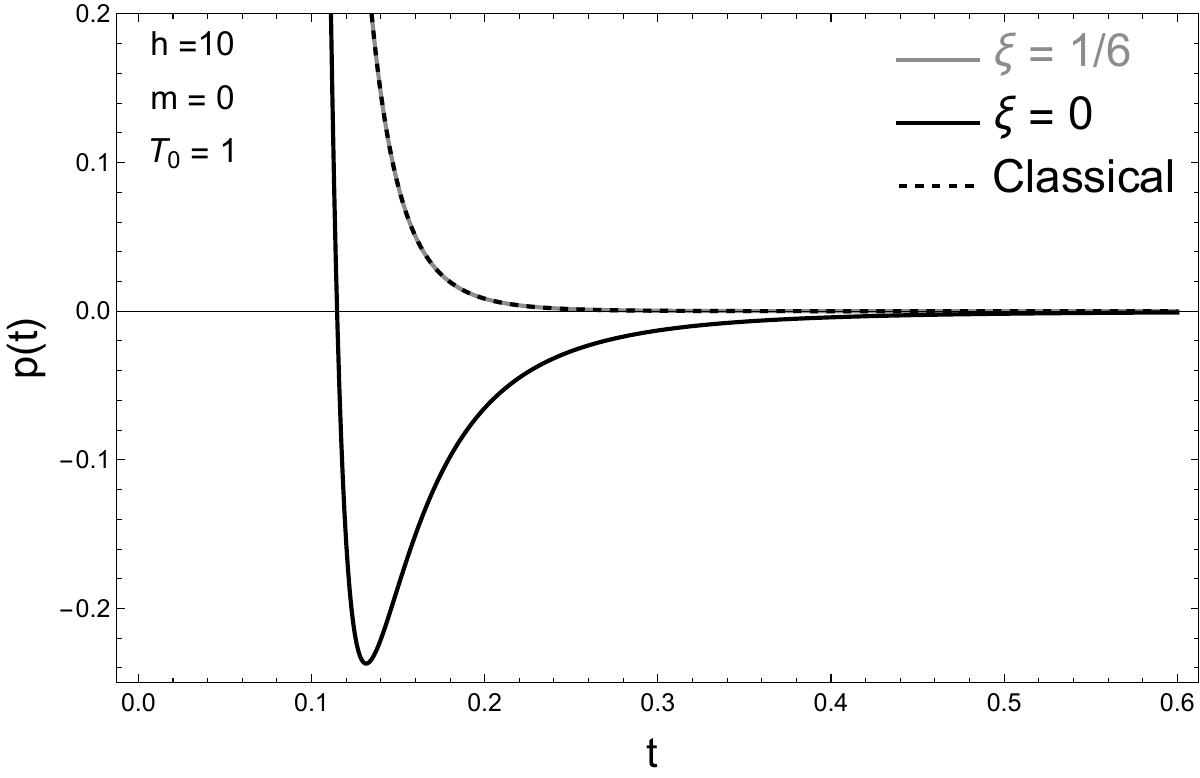}
  \caption{Pressure as a function of time for 
$a(t)=(t/t_0)^2$ for $h=10$ and $T_0=1$ (arbitrary units) for a massless field.}
  \label{zeromasspre}
\end{subfigure}
\caption{}
\end{figure}

For the late time limit $y \to 0$, one can use the expansion of Hankel functions of real 
order for small arguments and fixed order (see Appendix A, eq.~\eqref{HnuRsmall}), thereby 
obtaining the following asymptotic expression:
\begin{equation*}
    v_k \simeq -\sqrt{\dfrac{\pi}{4h}}\left(\dfrac{h}{\kk}\right)^\nu
    \dfrac{2^\nu \ii \Gamma(\nu)}{\pi}\dfrac{E_k}{y^{\nu-1/2}},
\end{equation*}
with $E_k \equiv C_k-D_k$, implying:
\begin{equation*}
    \begin{split}
 \left|v_k\right|^2&\sim\dfrac{1}{h}\left(\dfrac{h}{\kk}\right)^{2\nu}\dfrac{2^{2\nu-2}
 \left|\Gamma(\nu)\right|^2}{\pi}\dfrac{\left|E_k\right|^2}{y^{2\nu-1}},\\
\left|\dfrac{\di v_k}{\di y}\right|^2&\sim\dfrac{1}{h}\left(\dfrac{h}{\kk}\right)^{2\nu}
\dfrac{2^{2\nu-2}\left(\dfrac{1}{2}-\nu\right)^2\left|\Gamma(\nu)\right|^2}{\pi}
\dfrac{\left|E_k\right|^2}{y^{2\nu+1}},\\
\dfrac{\di v_k}{\di y}v^*_k&\sim\dfrac{1}{h}\left(\dfrac{h}{\kk}\right)^{2\nu}
\dfrac{2^{2\nu-2}\left(\dfrac{1}{2}-\nu\right)\left|\Gamma(\nu)\right|^2}{\pi}
\dfrac{\left|E_k\right|^2}{y^{2\nu}}.
    \end{split}
\end{equation*}
Plugging the above expressions in the equation \eqref{funzKGAMMA} we obtain the late time
approximation of the integrands of $\varepsilon$ and $p$:
\begin{equation}\label{IntegrandAsymptoticAccMassless}
    \begin{split}
\omega_kK_k&\sim\dfrac{h\left|E_k\right|^2\left|\Gamma(\nu)\right|^22^{2\nu-2}}{\pi}\left(\dfrac{h}{\kk}\right)^{2\nu}\times\begin{cases}
\dfrac{\mathrm{V}\left(\sigma,\xi\right)}{y^{2\nu+1}}&\xi\in\left(0,1/6\right),\\
\dfrac{\kk^2}{h^2}\dfrac{1}{y^{2\nu-1}},&\xi=0.
\end{cases},\\
\omega_k\Gamma_k&\sim-\dfrac{h\left|E_k\right|^2\left|\Gamma(\nu)\right|^22^{2\nu-2}}{\pi}\left(\dfrac{h}{\kk}\right)^{2\nu}\times\begin{cases}
\dfrac{\left|\mathrm{U}\left(\sigma,\xi\right)\right|}{y^{2\nu+1}}&\xi\in\left(0,1/6\right),\\
\dfrac{\kk^2}{3h^2}\dfrac{1}{y^{2\nu-1}},&\xi=0.
 \end{cases}
\end{split}
\end{equation}
where:
\begin{equation}\label{MasslessAccAsymptEoS}
   \begin{split}
 \mathrm{V}\left(\sigma,\xi\right)&=\left(\nu-\dfrac{1}{2}\right)^2+\left(1-6\xi\right)
 \left[\left(\dfrac{1}{2}-\nu\right)\dfrac{2\sigma}{\sigma-1}+\dfrac{\sigma^2}
 {\left(\sigma-1\right)^2}\right],\\
 \mathrm{U}\left(\sigma,\xi\right)&=\mathrm{V}\left(\sigma,\xi\right)-4\xi
 \left[\left(\nu-\dfrac{1}{2}\right)^2+\left(1-6\xi\right)\dfrac{\sigma\left(2\sigma-1\right)}
 {\left(\sigma-1\right)^2}\right]. 
   \end{split}
\end{equation}
From the eq. \eqref{IntegrandAsymptoticAccMassless} and taking into account that,
according to the definition \eqref{yeta}:
\begin{equation*}
    a^{\left(\sigma-1\right)/\sigma}=\frac{1}{y}
\end{equation*}
one finds that, for a massless field, $\varepsilon$ and $p$ decrease with the same power 
of the scale factor:
\begin{equation*}
    \varepsilon,\ p \propto a^r,
\end{equation*}
with $r$ given by:
\begin{equation}\label{rsigma}
    r\equiv\sqrt{1+\left(1-6\xi\right)\frac{4\sigma\left(2\sigma-1\right)}
    {\left(\sigma-1\right)^2}}\frac{\left(\sigma-1\right)}{\sigma}-4\mp1,
\end{equation}
where the minus applies for $\xi=0$ and the plus for $\xi\in\left(0,1/6\right)$.
For the minimal coupling $\xi=0$ it can be shown that:
\begin{equation*}
-3 < r < -2 \qquad\mbox{for}\ \sigma>1,
\end{equation*}
where the maximum value $-2$ is reached in the de Sitter limit $\sigma\to\infty$. 
For a non-adiabatic condition ($h=10,T_0=1$ in arbitrary units), in the minimal coupling, the 
energy density shows a significant deviation from its classical value and the pressure shows a 
negative dip before rising back and approaching zero according to $1/a^r$, with $r= -5/2$ for 
$\sigma=2$ according to the formula \eqref{rsigma} (see figure \ref{zeromassend} and figure 
\ref{zeromasspre}). 

According to the equation \eqref{IntegrandAsymptoticAccMassless}, the $\kk-$dependent term 
of the late time expansion for both $K_k$ and $\Gamma_k$ is the same, thus the resulting 
equation of state can be directly inferred from the ratio of the two integrand functions. 
For $\xi\in(0,1/6)$ we have:
\begin{equation}\label{pepsmassless}
    \lim_{t\to\infty}\dfrac{p}{\varepsilon}=-\dfrac{\left|\mathrm{U}\left(\sigma,\xi\right)\right|}
    {\mathrm{V}\left(\sigma,\xi\right)},\quad 0<\xi<\dfrac{1}{6}.
\end{equation}
and, for $\xi=0$:
\begin{equation*}
    \lim_{t\to\infty}\dfrac{p}{\varepsilon}= - \dfrac{1}{3},\qquad\qquad\forall\ \sigma>1,
\end{equation*}
which is precisely what we expected based on the arguments laid out in Section \ref{sec:gen}.
In addition, equation \eqref{MasslessAccAsymptEoS} shows that the functions $\mathrm{V}$ and
$\mathrm{U}$ are positive and negative, respectively, for any $\xi\in(0,1/6)$, so
that the asymptotic $w$ is always negative. The discontinuity in the asymptotic value of the 
equation of state for $\xi\to0^+$ is owing to $\xi$ coupling acting as an effective time dependent 
mass term, as it was mentioned in Section \ref{sec:intro}. Hence, while for $\xi=0$ there is no 
effective additional mass term and the dynamic of the modes implies a constant asymptotic value 
$p/\varepsilon = -1/3$, for $\xi\to0^+$ but $\xi\ne0$, the dynamic of the massless field is actually 
similar to that of a massive minimally coupled field, where the mass is provided by the $\xi$ term.
Indeed, the asymptotic ratio $p/\varepsilon$ depends on $\xi$ and $\sigma$ and can, according to
the equation \eqref{pepsmassless}, even attain the value $-1$ for $\sigma \to \infty$ like in 
the de Sitter case.
\begin{figure}[H]
\centering
\begin{subfigure}{.45\textwidth}
  \centering
  \includegraphics[width=.99\linewidth]{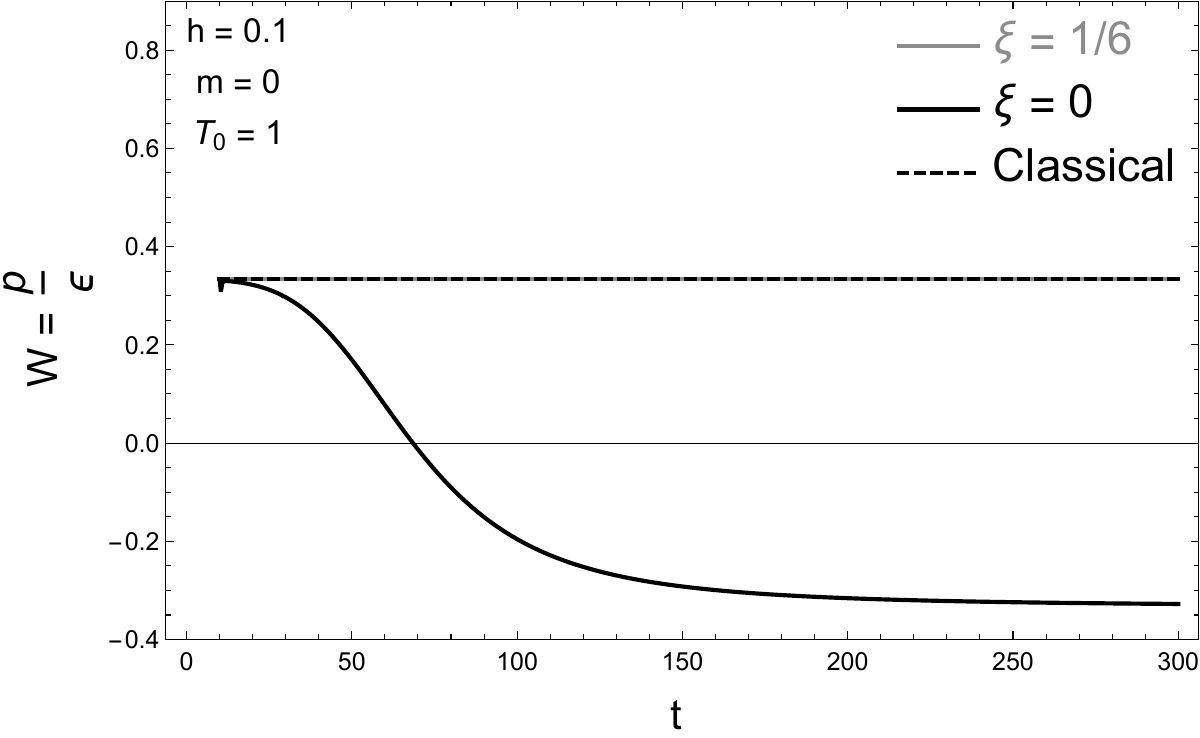}
  \caption{Ratio of pressure and energy density as a function of time for 
$a(t)=(t/t_0)^2$ for $h=0.1$ and $T_0=1$ (arbitrary units) for a massless field.}
  \label{zeromass3}
\end{subfigure}%
\hspace{1cm}
\begin{subfigure}{.45\textwidth}
  \centering
  \includegraphics[width=.99\linewidth]{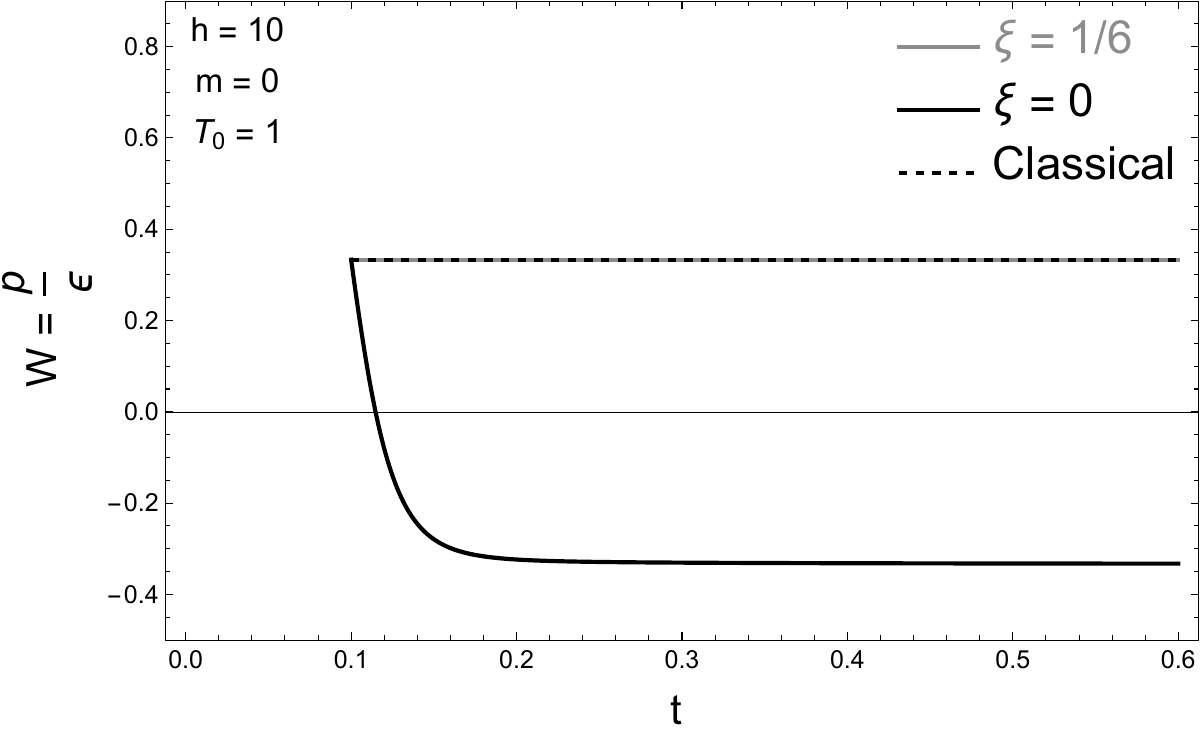}
  \caption{Ratio of pressure and energy density as a function of time for 
$a(t)=(t/t_0)^2$ for $h=10$ and $T_0=1$ (arbitrary units) for a massless field.}
  \label{zeromass4}
\end{subfigure}
\caption{}
\end{figure}
While the long-time limit of the equation of state is independent of the initial conditions, 
the time needed to get to this limiting value appears to depend, just like in the massive case, on 
whether the decoupling is adiabatic or non-adiabatic. This expectation is confirmed by the 
numerical computations of the ratio $w$ reported in figs.~\ref{zeromass3} and \ref{zeromass4}
for a ratio $h/T_0=0.1$ and $h/T_0=10$ respectively.

\section{Summary and discussion}
\label{sec:discu}

In this study of the equation of motion of the free scalar field in the cosmological FLRW 
background with several expansion laws of the scale factor $a(t)$ after the decoupling from
the thermal bath we have shown, as envisaged in Section \ref{sec:gen}, the existence of 
large quantum corrections to energy density and pressure if the field is in a non-adiabatic 
evolution regime, that is when the expansion rate is larger than the mass and/or 
the decoupling temperature. Such corrections can be large enough to modify their classical
dependences on the scale factor and to make the pressure largely negative. In some expansion 
scenarios, pressure attains its minimal allowed value (for the minimal coupling $\xi=0$), 
that is $-\varepsilon$ if $m\ne 0$ and $-\varepsilon/3$ if $m=0$.
Even for expansion rates which invariably imply an asymptotic adiabatic regime, there might be 
non-adiabatic transients which drive the pressure to large negative values before entering
the typical adiabatic oscillating regime with a frequency equal to the mass. 

In our calculations we used the approximation of a sudden decoupling at fixed time; in a 
more realistic study, one should take into account that different modes decouple over
a finite amount of time, thus at different temperatures of the plasma. Depending on the
magnitude of this time, some features will vanish and other will survive. For instance,
if the duration of the decoupling is comparable or larger than the inverse of the mass of the 
field, which sets the frequency of the oscillations of pressure, the oscillatory pattern of 
the pressure will vanish. Conversely, all the asymptotic $t\to \infty$ limits should be stable 
against a broadening of the freeze-out process.

As it has been mentioned, the purpose of this work has been to study the problem of quantum 
field effects on free-streaming in a cosmological background and not the actual cosmological
problem. Nevertheless, it is clear that the possibility of negative pressure at some stage may 
have important consequences on the evolution of the universe as a whole as well as on the structure 
formation \cite{Brandenberger:1995qk} because quantum correction to statistical fluctuations of energy 
density are also implied. Moreover, lot of Dark Matter candidates such as $\mathrm{WIMP}$ \cite{Arcadi:2017kky} 
or $\mathrm{SIMP}$ \cite{Hochberg:2014dra} and some light Dark Matter models \cite{Hannestad:2003ye,Hannestad:2005df} 
suggest that the current abundances are generated by a thermal decoupling \cite{Steigman:2012nb} 
thus corrections to the \emph{free-streaming} solution can alter the predictions and the bounds 
of the model. Of course, the study of the consequences requires specific assumptions on the models 
and the introduction of the back-reaction of the field on the metric, in other words the coupling of the 
Klein-Gordon with the cosmological Friedman equations. This will be the subject of further work.

\section*{Acknowledgments}

We are very grateful to Marko Simonovic for enlightening discussions and for 
reading the manuscript.



\appendix

\section{Hankel functions and their asymptotic expansions}
\label{app:hankel}

The Hankel functions can be defined in terms of a linear combination of the Bessel functions 
of the first and second kind: 
\begin{equation*}
    \mathrm{H}^{(1)}_\nu(z)=\mathrm{J}_\nu(z)+\ii \mathrm{Y}_\nu(z),\qquad
    \mathrm{H}^{(2)}_\nu(z)=\mathrm{J}_\nu(z)-\ii \mathrm{Y}_\nu(z),
\end{equation*}
The integral representation is \cite{Grad}:
\begin{equation}\label{HintegralRep}
\begin{split}
\huno(z)&=\dfrac{\e^{-\nu\pi \ii/2}}{\pi \ii}\int^{+\infty}_{-\infty}
\di \lambda \; \e^{\ii z\cosh\lambda-\nu\lambda}\\
\hdue(z)&=-\dfrac{\e^{\nu\pi \ii/2}}{\pi \ii}\int^{+\infty}_{-\infty}
\di \lambda \; \e^{-\ii z\cosh\lambda-\nu\lambda},
\end{split}
\end{equation}
where both $\nu$ and $z$ are, in general, complex numbers. From the above expressions it can be readily shown
that the two Hankel functions fulfill the relation:
\begin{equation}\label{HdueComplexConjugate}
    \hdue(z)^*=\mathrm{H}^{(1)}_{\nu^*}\left(z^*\right).
\end{equation}
For a generic order $\nu$ and argument $z$ the Hankel functions satisfy the following Wronskian 
condition \cite{Grad}:
\begin{equation}\label{WronskianoH}
    \hunopiu(z)\hdue(z)-\huno(z)\hduepiu(z)=-\dfrac{4i}{\pi z},
\end{equation}
where $\nu$ and $z$ can be both real or imaginary.
The derivative of the Hankel function respect to its argument can be related with the same 
Hankel function \cite{Grad}:
\begin{equation}\label{HrecursiveDeriv}
    \dfrac{\di\mathrm{H}^{(1,2)}_\nu(z)}{\di z}=
    \dfrac{\nu}{z}\mathrm{H}^{(1,2)}_\nu(z)-\mathrm{H}^{(1,2)}_{\nu+1}(z),\quad 
    \dfrac{\di\mathrm{H}^{(1,2)}_\nu(z)}{\di z}=-\dfrac{\nu}{z}\mathrm{H}^{(1,2)}_\nu(z)
     +\mathrm{H}^{(1,2)}_{\nu-1}(z),
\end{equation}
and
\begin{equation}\label{Hrecursivenupiuomenouno}
    \dfrac{2\nu}{z}\mathrm{H}^{(1,2)}_{\nu}(z)=\mathrm{H}^{(1,2)}_{\nu+1}(z)+\mathrm{H}^{(1,2)}_{\nu-1}(z).
\end{equation}

We are mostly interested in the behaviour of the Hankel functions for both small and large 
real arguments. The relevant expansions depend on the order $\nu$ of the function. We first 
consider the case of real order, $\nu\in\mathbb{R}$. From the equation \eqref{HdueComplexConjugate} 
we have:
\begin{equation*}
    \hdue(z)^*=\huno(z).
\end{equation*}
Then, from the power expansion of the Bessel functions, for small argument and positive order 
\cite{Grad}:
\begin{equation}\label{HnuRsmall}
    \hdue(z)\sim-\huno(z)\sim\dfrac{i}{\pi}\Gamma(\nu)\left(\dfrac{2}{z}\right)^\nu,
    \quad z\to0\quad \mbox{for}\ \nu>0.
\end{equation}
For small argument and negative order, since:
\begin{equation}\label{NegativeOrderHunHdue}
    \mathrm{H}^{(1)}_{-\nu}(z)=\e^{\ii\nu\pi }\mathrm{H}^{(1)}_\nu(z),
    \qquad \mathrm{H}^{(2)}_{-\nu}(z)=\e^{-\ii\nu\pi}\mathrm{H}^{(2)}_\nu(z),
\end{equation}
then
\begin{equation*}
    \mathrm{H}^{(2)}_{-\nu}(z)\sim\dfrac{\ii\e^{-\ii\nu\pi}}{\pi}\Gamma\left(\nu\right)
    \left(\dfrac{2}{z}\right)^\nu
\end{equation*}
For large arguments and fixed real order the following asymptotic form holds regardless 
of the sign of $\nu$:
\begin{equation}\label{HnuRhigh}
    \hdue(z)\sim\huno(z)^*\sim\sqrt{\dfrac{2}{\pi z}}\e^{-\ii \left(z-\nu\pi/2-\pi/4\right)},
    \quad z\gg1.
\end{equation}
For a pure imaginary order $\nu\equiv \ii \mu$ with real $\mu$, from the equations~\eqref{HintegralRep} 
and \eqref{NegativeOrderHunHdue} a relation between $\hunoC(z)$ and $\hdueC(z)$ can be obtained:
\begin{equation}\label{HunoHdueIM}
    \hunoC(z)^*=\e^{\mu\pi}\hdueC(z),\quad \hdueC(z)^*=\e^{-\mu\pi}\hunoC(z),
\end{equation}
which, combined with the \eqref{HrecursiveDeriv} and \eqref{Hrecursivenupiuomenouno} yields the 
useful identity:
\begin{equation*}
    \mathrm{H}^{(1,2)}_{i\mu-1}(z)=-\mathrm{H}^{(1,2)}_{i\mu+1}(z)+\dfrac{2i\mu}{z}
    \mathrm{H}^{(1,2)}_{i\mu}(z).
\end{equation*}
For the purpose of this work, we also need the expansions of Hankel functions of imaginary order 
with large $\mu$ and large argument \cite{Olver1954}:
\begin{equation}\label{HnuIhigh}
\hdueC\left(\mu x\right)\sim\sqrt{\dfrac{2}{\pi\mu}}\left(1+x^2\right)^{-1/4}
 \e^{-\ii\left(\mu\zeta(x)-\pi/4\right)-\mu\pi/2};\quad \mu\gg1 \quad x = {\cal O}(1).
\end{equation}
where the function $\zeta(x)$ reads:
\begin{equation}\label{zeta(x)}
    \zeta(x)=\sqrt{1+x^2}+\log\left(\dfrac{x}{1+\sqrt{1+x^2}}\right),
\end{equation}
and fulfills the following useful identity:
\begin{equation*}
    \dfrac{x}{\sqrt{1+x^2}}\dfrac{\di\zeta(x)}{\di x}=1
\end{equation*}
The expansion of $\hunoC$ in the same limit can be obtain using eq.~\eqref{HnuIhigh} and the 
relation \eqref{HunoHdueIM}.

Finally, since:
\begin{align*}
\e^{\ii\mu\zeta(x)}&=\exp\left[\ii \mu\sqrt{1+x^2}+\ii\mu\log\left(\dfrac{x}{1+\sqrt{1+x^2}}
\right)\right]
=\exp\left[\ii\mu\sqrt{1+x^2}+\log\left(\dfrac{x}{1+\sqrt{1+x^2}}\right)^{\ii \mu}\right]\\
&=\e^{\ii\mu\sqrt{1+x^2}}\left(\dfrac{x}{1+\sqrt{1+x^2}}\right)^{\ii\mu}
\end{align*}
we have, in the limit $x \to 0$:
\begin{equation*}
\e^{\ii \mu\zeta(x)}=\e^{\ii \mu\sqrt{1+x^2}}\left(\dfrac{x}{1+\sqrt{1+x^2}}\right)^{\ii \mu}
\simeq \e^{\ii \mu} \left(\dfrac{x}{2}\right)^{\ii\mu} = \e^{\ii \mu \log(ex/2)}
\end{equation*}
and in the same limit the \eqref{HnuIhigh} becomes:
\begin{equation}\label{HnuIsmall}
    \hdueC(\mu x) \simeq \sqrt{\dfrac{2}{\pi\mu}}
    \e^{-\mu\pi/2}\exp\left\{-\ii \left[\log\left(\dfrac{\e x}{2}\right)-\dfrac{\pi}{4}\right]\right\}
\end{equation}
The similar expansion of $\hunoC$ can be obtained by using the eq.~\eqref{HnuIhigh}.

\section{Adiabatic decoupling}
\label{app:adiadec}

Whenever the adiabatic parameter $\mathcal{A}_k$ in the equation \eqref{adiapar} is $\ll 1$ 
the solution to the equation of motion is well approximated by a combination of plane waves \eqref{adia}:
\begin{equation*}
    v_k(\eta)=A_k\dfrac{\e^{-\ii\int_\eta\Omega_k}}{\sqrt{2\Omega_k(\eta)}}+B_k\dfrac{\e^{\ii\int_\eta\Omega_k}}{\sqrt{2\Omega_k(\eta)}},
\end{equation*}
If the decoupling occurs in an adiabatic regime, that is when $\mathcal{A}_k(0) \ll 1$ 
(see equation \eqref{adzero}) we can apply the initial conditions \eqref{condin} to get an 
approximated expression for the coefficients $A_k$ and $B_k$. We also require that the higher
order derivatives appearing in the equation \eqref{adzero} are negligible with respect to
the mass scale, so that $\mathcal{A}_k$ is well approximated by:
\be\label{adzero2}
\mathcal{A}_k(0) \simeq H(0) \frac{m^2}{\epsilon_k^3}
\ee
with $\epsilon_k = \sqrt{\kk^2+m^2}$.

By taking the derivative of the mode function $v_k$:
\begin{equation*}
    v'_k(\eta)=\sqrt{\dfrac{\Omega_k(\eta)}{2}}\left[A_k\e^{-\ii \int_\eta\Omega_k}
    \left(-\ii-\dfrac{\mathcal{A}_k(\eta)}{2}\right)
    +B_k \e^{\ii \int_\eta\Omega_k}\left(\ii-\dfrac{\mathcal{A}_k(\eta)}{2}\right)\right].
\end{equation*}
where the definition \eqref{adiapar} for $\mathcal{A}_k$ has been used, and enforcing the 
initial conditions \eqref{condin} we obtain the following equations:
\begin{equation*}
\begin{split}
\dfrac{A_k}{\sqrt{2\Omega_k(0)}}+\dfrac{B_k}{\sqrt{2\Omega_k(0)}}&=
\dfrac{1}{\sqrt{2\omega_\xi\left(0,\kk\right)}},\\
A_k\left(-\ii-\dfrac{\mathcal{A}_k(0)}{2}\right)+B_k\left(\ii-\dfrac{\mathcal{A}_k(0)}{2}\right)
&=-\ii \sqrt{\dfrac{\omega_\xi\left(0,\kk\right)}{\Omega_k(0)}}+\dfrac{\left(1-6\xi\right)a'(0)}
{\sqrt{\Omega_k(0)\omega_\xi\left(0,\kk\right)}}.
\end{split}
\end{equation*}
whose solution is:
\begin{equation}\label{ab}
    \begin{split}
A_k&=\dfrac{1}{2}\sqrt{\dfrac{\omega_\xi\left(0,\kk\right)}
{\Omega_k(0)}}-\dfrac{i}{2}\sqrt{\dfrac{\Omega_k(0)}{\omega_\xi(0)}}
\left(\ii+\dfrac{\mathcal{A}_k(0)}{2}\right)-\dfrac{i}{2}\dfrac{\left(1-6\xi\right)
a'(0)}{\sqrt{\Omega_k(0)\omega_\xi\left(0,\kk\right)}},\\
B_k&=-\dfrac{1}{2}\sqrt{\dfrac{\omega_\xi\left(0,\kk\right)}
{\Omega_k(0)}}-\dfrac{\ii}{2}\sqrt{\dfrac{\Omega_k(0)}{\omega_\xi(0)}}
\left(\ii+\dfrac{\mathcal{A}_k(0)}{2}\right)-\dfrac{\ii}{2}\dfrac{\left(1-6\xi\right)
a'(0)}{\sqrt{\Omega_k(0)\omega_\xi\left(0,\kk\right)}}.
    \end{split}
\end{equation}
For we have supposed that higher order derivatives of the scale factor $a(\eta)$ are
negligible compared to the mass, we can approximate both $\Omega_k(0)$ in eq. \eqref{Omega}
and $\omega_\xi\left(0,\kk\right)$ in eq. \eqref{omegaxi} with:
\begin{equation*}
    \Omega_k(0)\simeq\omega_\xi\left(0,\kk\right)\simeq\sqrt{\kk^2+m^2} = \epsilon_k
\end{equation*}
hence the coefficients in \eqref{ab} can be approximated, for $\kk \ll m$ as:
\begin{equation}
    \begin{split}
        A_k&\simeq 1-\dfrac{i\left(3-12\xi\right)}{4}{\mathcal A}_k(0) ,\\
        B_k&\simeq -\dfrac{i\left(3-12\xi\right)}{4}{\mathcal A}_k(0) 
    \end{split}
\end{equation}

\end{document}